\makeatletter\let\ifGm@compatii\relax\makeatother
\documentclass{elsarticle}

\usepackage{latexsym}
\usepackage{amssymb}
\usepackage{amsmath}
\usepackage{amsthm}
\usepackage{amsfonts}
\usepackage{mathabx}

\usepackage{autobreak}
\usepackage{graphicx}

\newtheorem{thm}{Theorem}
\newtheorem{lem}{Lemma}
\newtheorem{pn}{Proposition}
\newtheorem{cor}{Corollary}

\newproof{pf}{Proof}

\newdefinition{dn}{Definition}
\newdefinition{rmk}{Remark}

\newenvironment{claim}{\begin{list}{}{\listparindent\parindent%
                                      \leftmargin0cm\parsep\parskip}%
                                      \item[] \refstepcounter{claim}%
                                      {\em Claim \arabic{claim} }}%
                                      {\end{list}}

\newcounter{claim}

                                      {\end{list}}

\newcounter{case}[lem]

\newenvironment{case}{\begin{list}{}{\listparindent\parindent%
                                      \leftmargin0cm\parsep\parskip}%
                                      \item[] \refstepcounter{case}%
                                      {\em Case \arabic{case} }}%
                                      {\end{list}}
                                      
\newcounter{subcase}[lem]

                                      {\end{list}}

                                      {\end{list}}                                             

\def\hat{\widehat}
\def\tilde{\widetilde}

\newcommand{\fsubset}{\subseteq_\mathrm{fin}}
\newcommand{\appmap}[3]{\mbox{$#1 \colon #2 \trianglelefteq #3$}}
\newcommand{\appmapf}[3]{\mbox{$#1 \colon #2 \varpropto #3$}}
\newcommand{\cormap}[3]{\mbox{$#1 \colon #2 \bowtie #3$}}

\newcommand{\bA}{\mathbb{A}}

\newcommand{\GG}{\mathbb{G}}
\newcommand{\HH}{\mathbb{H}}

\newcommand{\DD}{\mathbb{D}}

\newcommand{\PP}{\mathbb{P}}

\newcommand{\bS}{\mathbb{S}}

\newcommand{\CCC}{\mathcal{C}}

\newcommand{\EEE}{\mathcal{E}}
\newcommand{\FFF}{\mathcal{F}}

\newcommand{\KKK}{\mathcal{K}}

\newcommand{\LLL}{\mathcal{L}}

\newcommand{\SSS}{\mathcal{S}}
\newcommand{\TTT}{\mathcal{T}}
\newcommand{\UUU}{\mathcal{U}}
\newcommand{\VVV}{\mathcal{V}}
\newcommand{\WWW}{\mathcal{W}}

\newcommand{\bC}{\mathbf{C}}
\newcommand{\FF}{\mathfrak{F}}

\newcommand{\KF}{\mathfrak{K}}

\newcommand{\UF}{\mathfrak{U}}
\newcommand{\VF}{\mathfrak{V}}
\newcommand{\XF}{\mathfrak{X}}
\newcommand{\YF}{\mathfrak{Y}}
\newcommand{\ZF}{\mathfrak{Z}}

\newcommand{\IR}{\mathrm{IR}}

\newcommand{\powf}[1]{\mathcal{P}_{\mathrm{fin}}(#1)}
\newcommand{\dda}{\mathord{\mbox{\makebox[0pt][l]{\raisebox{-.4ex}
                           {$\downarrow$}}$\downarrow$\,}}}
\newcommand{\tdash}{\mathrel{\tilde{\vDash}}}
\newcommand{\odash}{\mathrel{\ov{\vDash}}}

\newcommand{\DOM}{\mathbf{DOM}}
\newcommand{\AB}{\mathbf{AB}}
\newcommand{\SCIS}{\mathbf{SCIS}}
\newcommand{\CIS}{\mathbf{CIS}}
\newcommand{\CIF}{\mathbf{CIF}}
\newcommand{\SIF}{\mathbf{SIF}}
\newcommand{\CSL}{\mathbf{CSL}}

\newcommand{\set}[2]{\mbox{$\{\,#1 \mid #2 \,\}$}}
\newcommand{\fun}[3]{\mbox{$#1 \colon #2 \rightarrow #3$}}

\newcommand{\bt}{\mathbf{t}}

\makeatletter
\newlength{\doublefracgap}
\DeclareRobustCommand{\doublefrac}[2]{%
  \mathinner{\mathpalette\doublefrac@{{#1}{#2}}}%
}
\newcommand{\doublefrac@}[2]{\doublefrac@@#1#2}
\newcommand{\doublefrac@@}[3]{%
  \ooalign{%
    \raisebox{\doublefracgap}{$\m@th#1\frac{#2}{\phantom{#3}}$}\cr
    \raisebox{-\doublefracgap}{$\m@th#1\frac{\phantom{#2}}{#3}$}\cr
  }%
}
\newcommand{\ddoublefrac}[2]{{\displaystyle\doublefrac{#1}{#2}}}

\makeatother

\newcommand{\ov}[1]{\overline{#1}}

\def\con{\mathop{\mathstrut\rm Con}\nolimits}
\def\Con{\mathop{\mathstrut\rm CON}\nolimits}
\def\kand{\mathop{\mathrel{\&}}}
\def\kor{\mathop{\mathrel{\rm or \,}}}

\def\mmodels{\mathrel {||}\joinrel \Relbar}

\newcommand{\corel}{\mathrel\arrowvert\!\sim}

\def\int{\mathop{\mathstrut\rm int}\nolimits}

\def\Id{\mathop{\mathstrut\rm Id}\nolimits}

\def\ID{\mathop{\mathstrut\mathcal{I}}}

\begin{document}

\begin{frontmatter}

\title{Domains, Information Frames, and Their Logic\tnoteref{t1}}
\tnotetext[t1]{This research was funded by DFG grant no.\ 549144494.}

\author{Dieter Spreen\corref{cor1}}
\ead{spreen@math.uni-siegen.de}

\cortext[cor1]{Corresponding author}

\affiliation{organization={Department of Mathematics, University of Siegen},
postcode={57068},
postcodesep={},
city={Siegen},
country={Germany}}

\begin{abstract}
In \cite{sp25}, continuous information frames were introduced that capture exactly all continuous domains. They are obtained from the information frames considered in \cite{sp21} by omitting the conservativity requirement. Information frames generalise Scott's information systems~\cite{sc82}: Instead of the global consistency predicate, there  is now a local consistency predicate for each token. Strong information frames are obtained by strengthening the conditions for these predicates. Let $\CIF$ and $\SIF$ be the corresponding categories.

In \cite{sxx08} another generalisation of Scott's information systems was introduced which also exactly captures all continuous domains. As  shown in \cite{hzl15}, the definition can be simplified while maintaining the representation result. Let $\CIS$ and $\SCIS$ be the corresponding categories. It is shown that all these categories are equivalent. Moreover, the equivalence extends to the subcategories of (strong) continuous information frames with truth elements. Such information frames capture exactly all pointed continuous domains.

Continuous information frames are families of rudimentary logics, associated with each token is a local consistency predicate and an entailment relation. However, they lack the expressive power of propositional logic. In an attempt to make each of this logics more expressible, continuous stratified conjunctive logics are introduced. These are families of conjunctive logics. The category $\CSL$ of such logics is shown to be isomorphic to $\SIF_{\bt}$, the category of strong continuous information frames with a truth element.
\end{abstract}


\end{frontmatter}

\section{Introduction}\label{sec-intro}

Domain theory, introduced independently by Dana Scott~\cite{sc72} and Yuri L.\ Ershov~\cite{er74}, is about approximation, in particular about the approximation of ideal elements by already constructed ones. Intuitively, each domain element is thought of carrying a piece of consistent information. We will assume that this amount of information is closed under taking conclusions.

If $D$ is a domain and $X \subseteq D$, then the information pieces coming with the various elements of $X$ need not be consistent with each other, and of course the sum of these information pieces is not closed under conclusions. However, if $X$ has an upper bound $z$, then $z$ \emph{witnesses} that the cumulative information contained in $X$  is consistent, as this information is part of the information coming with $z$, and since the latter amount is closed under taking conclusions, it also contains the conclusions that can be drawn from the information coming with $X$. 
If $D$ is bounded-complete, each bounded subset has unique minimal such witness, which is not true for domains in general.

To formalise these ideas, Scott~\cite{sc72}  introduced the notion of an \emph{information system} $(A, \con, \vDash)$, where $A$ is a set of tokens (information pieces), $\con \subseteq \powf{A}$ a consistency predicate, and $\vDash \subseteq \con \times A$ an entailment relation. Under a few natural requirements on $A$, Scott showed that the consistent entailment-closed subsets of $A$ form a bounded-complete algebraic domain, and that up to isomorphism every bounded-complete algebraic domain can be obtained in this way. By slightly modifying the assumptions, Hoofman~\cite{ho93}  was able to extend the result to the case of  bounded-complete continuous domains. 

In  joint work with Luoshan Xu and Xuxin Mao~\cite{sxx08} another set of natural requirements was presented so that the information systems satisfying these requirements ---called \emph{continuous information systems}---capture exactly the continuous domains. Huang, Zhou and Li~\cite{hzl15} later showed that one of the axioms in this definition can be omitted. Accordingly, they called their information systems \emph{simplified continuous information systems}. As will be shown in this paper, the categories $\CIS$ and $\SCIS$ of continuous and simplified continuous information systems, respectively, are equivalent.

The proof of the representation theorem in \cite{sxx08} is an adaption of Scott's original construction. The main difference lies in the definition of the consistency predicate. In the general continuous case, bounded subsets $X \subseteq D$ need not have a least upper bound.  So, the essential idea in the modified definition of $\con$ was to consider only those finite subsets $X$ (of base elements) of $D$ that contain a maximal element (with respect to the order of approximation). 

Another, less restrictive approach would be  a formalism in which the consistency witnesses are made explicit.  Information frames~\cite{sp21} are obtained by following this idea. As shown, they exactly capture all L-domains. In this article, we slightly modify this notion and omit the conservativity requirement, resulting in the notion of a \emph{continuous information frame}.

At the \emph{International Symposium on Domain Theory (ISDT 2024)} Prof.\ Luoshan Xu posed the question of the relationship between continuous information frames and continuous information systems: Does every continuous information system lead to a continuous information frame and vice versa? Investigating this question was one of the main motivations for the present work. 

As will be shown, the category $\CIF$  of continuous information frames  is equivalent to the category $\SCIS$ of simplified continuous information systems, the latter of which is known to be equivalent to the category of  abstract bases and hence to the category of continuous domains~\cite{sxx08,hzl15}. 

A continuous information frame is similar to a Kripke frame. It consists of a set $P$ of tokens (atomic propositions) and an interpolative (in particular, dense) transitive binary relation $R$ on $P$. Associated with each node $p \in P$ is an information system $(P, \con_{p}, \Vdash_{p})$, where the sets in  $\con_{p}$ can be thought of as being those finite sets of atomic propositions that are consistent with proposition $p$, and $\Vdash_{p} \subseteq  \con_{p} \times P$ is an entailment relation for the (rudimentary) logic determined by proposition $p$. These relations are required to be sound, and closed under the rules Weakening, Interpolation and Cut. Consistency and entailment are inherited upwards along the relation $R$. Both families must also satisfy a global interpolation condition, which, in particular, allows the reversal of the cut rule.
The conditions to be fulfilled are such that each of the local information systems satisfies the conditions for a continuous information system in the sense of Hoofman~\cite{ho93}.

By what has been said above it follows that a continuous information frame is a family of rudimentary local logics. We will make this explicit by extending each frame to a family of conjunctive fragments of propositional sequent calculus, called \emph{continuous stratified conjunctive logic}: For every $p \in P$, the information system $(P_{p}, \con_{p}, \Vdash_{p})$ associated with frame node $p$ is related to a conjunctive sequent calculus $(\LLL(P_{p}), \vdash^{p})$. Locally, for  each $p$, the set of derivable sequents is closed under the usual rules known from elementary proof theory. In addition, the following global conditions hold for $p, q \in P$,  
\begin{itemize}

\item If $q \in P_{p}$ then $p \vdash^{p} q$.

\item If $p \vdash^{p} q$, then $P_{q} \subseteq P_{p}$ and for $\varphi \in \LLL(P_{q})$ and $\Gamma \fsubset \LLL(P_{q})$ with $\Gamma \vdash^{q} \varphi$, $\Gamma \vdash^{p} \varphi$.

\item For all $\varphi \in \LLL(P_{q})$ and $\Gamma \fsubset \LLL(P_{q})$ with $\Gamma \vdash^{q} \varphi$  there is some $r \in P_{q}$ so that $\Gamma \vdash^{q} r$ and $r \vdash^{r} \varphi$.

\end{itemize}
Similar families of conjunctive sequent calculi have been studied by Wang and Li~\cite{wl24} to characterise L-domains.

Appropriate morphisms will be introduced for continuous information frames and continuous stratified conjunctive logics, respectively; and it will be shown that the corresponding categories $\CIF$ and $\CSL$ are equivalent.

The paper is organised as follows: In Section~\ref{sec-basic} basic definitions and results from domain theory are presented. Definitions and first results on the information structures investigated in this paper and their morphisms can be found in Section~\ref{sec-infr}. 

Subsequently, in Section~\ref{sec-sysvsfr} the equivalence of the categories $\CIS$ and $\SCIS$ of continuous and simplified continuous information systems is derived. Moreover, it is shown that $\CIS$ is equivalent to the category $\SIF$ of strong continuous information frames. The strength condition is based on the idea that the consistency of the finite sets of atomic propositions  consistent with proposition $p$ should rest on the fact that their elements are entailed by $p$.

Strong continuous information frames are examined in more detail in Section~\ref{sec-strong}. The category $\SIF$ is a reflexive full subcategory of the category $\CIF$ of continuous information frames. It will be shown that both categories are equivalent and that this equivalence extends to their subcategories $\CIF_{\bt}$ and $\SIF_{\bt}$ of continuous and strong continuous information frames, respectively, with truth elements. 

Continuous information frames with truth elements  represent  exactly pointed continuous domains. They are examined in Section~\ref{sec-truth}, where it is shown that any continuous information frame can be extended by a truth element. As a consequence we obtain that the categories $\CIF$ and $\CIF_{\bt}$ are equivalent. 

Continuous stratified conjunctive logics and their morphisms, global consequence relations, are introduced in Section~\ref{sec-log}, and in Section~\ref{sec-inflog} it will be shown that their category $\CSL$ is isomorphic to the category  $\SIF_{\bt}$.  As a consequence of the results of this work, all categories examined here prove to be equivalent.

The paper finishes with a Conclusion. Initial results of this research were presented in an invited talk at the International Symposium on Domain Theory ISDT 2024.

\section{Domains: basic definitions and results}\label{sec-basic}
 
For any set $A$, we write $X \fsubset A$ to mean that $X$ is finite subset of $A$. The collection of all subsets of $A$ will be denoted by $\mathcal{P}(A)$ and that of all finite subsets by $\mathcal{P}_f(A)$. $\|A\|$ denotes the cardinality of $A$.

Let $\DD = (D, \sqsubseteq)$ be a poset. $\DD$ is \emph{pointed} if it contains a least element $\bot$.  A subset $S$ of $D$ is \emph{directed}, if it is non-empty and every pair of elements in $S$ has an upper bound in $S$. $\DD$ is a \emph{directed-complete partial order} (\emph{dcpo}), if every directed subset $S$ of $D$ has a least upper bound $\bigsqcup S$ in $D$.

Assume that $x, y$ are elements of a poset $\DD$. Then $x$ is said to \emph{approximate} $y$, written $x \ll y$, if for any directed subset $S$ of $D$ the least upper bound of which exists in $D$, the relation $y \sqsubseteq \bigsqcup S$ always implies the existence of some $u \in S$ with $x \sqsubseteq u$.   A subset $B$ of $D$ is a \emph{basis} of $\DD$, if for each $x \in D$ the set $\dda\!_B x = \set{u \in B}{u \ll x}$ contains a directed subset with least upper bound $x$; and a  directed-complete partial order $\DD$ is said to be \emph{continuous} (or a \emph{domain}) if it has a basis. Standard references for domain theory and its applications are~\cite{aj94,sh94,ac98,gh03}.

\begin{lem}\label{lem-preordprop}
In a poset $\DD$ the following statements hold for all $u, x, y, z \in D$: \begin{enumerate}
\item\label{lem-preordprop-0} The approximation relation $\ll$ is transitive.
\item\label{lem-preordprop-1} $x \ll y \Rightarrow x \sqsubseteq y$.
\item\label{lem-preordprop-2} $u \sqsubseteq x \ll y \sqsubseteq z \Rightarrow u \ll z$.
\item\label{lem_preordprop-5} If $\DD$ is a domain with basis $B$, and $M \fsubset D$, then
\begin{equation*}
M \ll x \Rightarrow (\exists v \in B) M \ll v \ll x,
\end{equation*}
where $M \ll x$ means that $m \ll x$, for any $m \in M$.

\end{enumerate}
\end{lem}

Property~\ref{lem_preordprop-5} is known as the \emph{interpolation law}. A set $B$ together with a transitive relation $\prec$ on $B$ satisfying the interpolation law is called \emph{abstract basis}. The basis of a continuous domain together with the approximation relation is a prime example of an abstract basis. 
Recall that for a transitive relation $\prec$ on a set $B$, a subset $S$ of $B$ is called \emph{dense} if for all $u, v \in B$ with $u \prec v$ there is some $w \in S$ such that $u \prec w \prec v$. If $B$ is dense, one also says that the relation $\prec$ is dense. For abstract bases $(B, \prec)$ we therefore have that $\prec$ is a dense transitive relation.

The usual morphisms between abstract bases are approximable relations.

\begin{dn}\label{dn-apprel}
 A relation $R$ between abstract bases $B$ and $C$ is called \emph{approximable} if the following conditions hold for all $u, u' \in B$ and $v, v' \in C$ and all finite subsets $M$ of $C$:
\begin{enumerate}

\item $(uRv \kand v \succ_{C} v') \Rightarrow u R v'$,
\item $(\forall v'' \in M)\, uRv'' \Rightarrow (\exists w \in C)\, (uRw \kand w \succ_{C} M)$,
\item $(u' \succ_{B} u \kand uRv) \Rightarrow u'Rv$,
\item $uRv \Rightarrow (\exists w \in B)\, (u  \succ_{B} w \kand wRv)$.

\end{enumerate}
\end{dn}

Obviously, for every abstract basis $(B, \prec)$, the relation $\prec$ is approximable. It is the identity morphism $\IR_{B}$ on $B$. The morphisms usually considered in case of continuous domains are Scott continuous functions.

\begin{dn}
Let $\DD$ and $\DD'$ be posets. A function $\fun{f}{D}{D'}$ is \emph{Scott continuous} if it is monotone and for any directed subset $S$ of $D$ with existing least upper bound,
\[
\bigsqcup f(S) = f(\bigsqcup S).
\]
\end{dn}

\begin{thm}
The category $\AB$ of abstract bases and approximable relations is equivalent to the category $\DOM$ of continuous
domains and Scott continuous functions.
\end{thm}

\section{Information structures and their morphisms}\label{sec-infr}

In order to give a characterisation of domains in the style of Scott's information systems~\cite{sc82} the present author in collaboration with L.\ Xu and X.\ Mao~\cite{sxx08} introduced continuous information systems.

\begin{dn}\label{dn-infosys}
Let $S$ be a set, $\con$ a collection of finite subsets of $S$ and $\Vdash \subseteq \con \times S$. Then $\bS = (S, \con, \Vdash)$ is a \emph{continuous information system} if the following conditions hold for all sets $X, Y \in \con$, elements $a \in S$ and finite subsets $F$ of $S$:
\begin{enumerate}

\item\label{dn-infosys-1} $\{a\} \in \con$,

\item\label{dn-infosys-2} $X \Vdash a \Rightarrow  X \cup \{a\} \in \con$,

\item\label{dn-infosys-3} $(Y \supseteq X \kand X \Vdash a) \Rightarrow Y \Vdash a$,

and, defining $X \Vdash Y$ to mean that $X \Vdash b$, for all $b \in Y$,

\item\label{dn-infosys-4} $(X \Vdash Y \kand Y \Vdash a) \Rightarrow X \Vdash a$,

\item\label{dn-infosys-5} $X \Vdash a \Rightarrow  (\exists Z \in \con)\, (X \Vdash Z \kand Z \Vdash a)$,

\item\label{dn-infosys-6} $X \Vdash F \Rightarrow (\exists Z \in  \con)\, (Z \supseteq F \kand X \Vdash Z)$.

\end{enumerate}
\end{dn}

If $(S, \con, \Vdash)$ is a continuous information system then the elements of $S$ are usually called \emph{tokens}, the sets in $\con$ \emph{consistent}
and the relation $\Vdash$ \emph{entailment relation}. Tokens should be thought of as atomic propositions giving information about data and consistent sets as representing consistent finite conjunctions of such propositions. The entailment relation then tells us which propositions are derivable from what.

In \cite{sxx08} it was shown that continuous information systems allow the generation of all abstract bases. As was observed by Huang, Zhou and Li~\cite{hzl15} later on, Condition~\eqref{dn-infosys-2} is not used in the derivation of this result. They called systems $(S, \con, \Vdash)$ which are only required to satisfy Condition~\eqref{dn-infosys-1} and Conditions~\eqref{dn-infosys-3}-\eqref{dn-infosys-6} \emph{simplified continuous information systems}.

The usual morphisms between simplified continuous information systems are approximable mappings.

\begin{dn}\label{dn-appm}
An \emph{approximable mapping} $H$ between simplified continuous information system $\bS$ and $\bS'$, written $\appmap{H}{\bS}{\bS'}$,  is a relation between $\Con$ and $S'$ satisfying for all $X, X' \in \Con$, $Y \in \Con'$ and $b \in S'$, as well as all finite subsets $F$ of $S'$ the following  conditions:
\begin{enumerate}
\item\label{dn-appm-1}
$(XHY \kand Y \Vdash' b) \Rightarrow XHb$,

\item\label{dn-appm-2}
$(X \supseteq X' \kand X' Hb) \Rightarrow X Hb$,

\item\label{dn-appm-3} 
$(X \Vdash X' \kand X' Hb) \Rightarrow XHb$,

\item\label{dn-appm-4}
$XHb \Rightarrow (\exists Z \in \Con) (\exists Z' \in \Con')\, (X \Vdash Z \kand  ZHZ' \kand Z' \Vdash'  b)$,

\item\label{dn-appm-5}
$XHF \Rightarrow (\exists Z \in \Con')\, (Z \supseteq F \kand XHZ)$,

\end{enumerate}
where $XHF$ means that for all $a \in F$,  $XHa$.
\end{dn}

In applications it is sometimes preferable to have Condition~\eqref{dn-appm-4} split up into two conditions which state interpolation for the domain and the range of the approximable mapping, separately.

\begin{lem}\label{lem-amintan}
Let $\bS$ and $\bS'$ be simplified continuous information systems. Then, for any $X \in \Con$, $F \fsubset S'$ and $H \subseteq \Con \times S'$ satisfying Condition~\ref{dn-appm}\eqref{dn-appm-5} , Condition~\ref{dn-appm}\eqref{dn-appm-4} is equivalent to the following Conditions~\eqref{lem-amintan-1} and \eqref{lem-amintan-2}:
\begin{enumerate}

\item\label{lem-amintan-1}
$X H F \Rightarrow (\exists Z \in \Con)\, X \Vdash Z \kand Z H F$, 

\item\label{lem-amintan-2}
$X H F \Rightarrow (\exists Z' \in \Con')\, X H Z' \kand Z' \Vdash' F$.

\end{enumerate}
\end{lem}

As is readily seen, the entailment relation in a simplified continuous information system $\bS$ is an approximable mapping. It is the identity $\Id_{\bS}$ on $\bS$. The composition $H \circ G$ of two approximable mappings $\appmap{H}{\bS}{\bS'}$ and $\appmap{G}{\bS'}{\bS''}$  is defined as the usual composition of binary relations.

In what follows let $\SCIS$ be the category of simplified continuous information systems and approximable mappings, and $\CIS$ be the full subcategory of continuous information systems.

The requirements on the consistency predicate are very weak compared with the usual consistency notion in logic. An obvious property of the latter is that every subset of a consistent set of propositions is consistent as well. A further drawback of the above notion of continuous information systems is that the consistency witnesses are not visible or accessible. To resolve these issues information frames were introduced by the present author~\cite{sp21,sp25}.  

\begin{dn}\label{dn-infofr}
Let $A$ be a set,  $(\con_i)_{i \in A}$ be a family of subsets of $\mathcal{P}_f(A)$, and $(\vDash_i)_{i \in A}$ be a family of relations $\vDash_i  \subseteq \con_i \times A$. For $i, j \in A$
set
\begin{equation*}
iRj \Leftrightarrow \{ i \} \in \con_{j}.
\end{equation*}
Then $\bA = (A, (\con_i)_{i \in A}, (\vDash_i)_{i \in A})$ is a \emph{continuous information frame} if the following conditions hold, for all $a \in A$ and all finite subsets $X, Y$ of $A$:
\begin{enumerate}
\item Local conditions, for every $i \in A$:

\begin{enumerate}
\item\label{dn-infofr-1}
$\{i\} \in \con_i$ \hfill (self consistency),

\item\label{dn-infofr-2}
$Y \subseteq X \kand X \in \con_i \Rightarrow Y \in \con_i$ \hfill (consistency preservation),

\hspace{-\leftmargin}
and, defining $X \vDash_i Y$ to mean that $X \vDash_i b$, for all $b \in Y$,

\item\label{dn-infofr-4}
$X \in \con_{i}\mbox{} \kand X \vDash_i Y \Rightarrow Y \in \con_i$ \hfill (soundness),

\item\label{dn-infofr-5}
$X, Y \in \con_i\mbox{} \kand Y \supseteq X \kand X \vDash_i a \Rightarrow Y \vDash_i a$ \hfill (weakening),

\item\label{dn-infofr-6}
$X \in \con_i\mbox{} \kand X \vDash_i Y \kand Y \vDash_i a \Rightarrow X \vDash_i a$ \hfill (cut),

\end{enumerate}

\item Global conditions, for all $i, j \in A$

\begin{enumerate}

\item\label{dn-infofr-7}
$i R j \Rightarrow \con_i \subseteq \con_j$ \hfill (consistency transfer),

\item\label{dn-infofr-9}
$i R j \kand X \in \con_i \mbox{} \kand X \vDash_i a \Rightarrow X \vDash_j a$ \hfill (entailment transfer),

\item\label{dn-infofr-11}
$X \vDash_i Y \Rightarrow (\exists e \in A) (\exists Z \in \con_{e}) X \vDash_i (\{ e \} \cup Z) \kand Z \vDash_{e} Y$ \hfill (interpolation).

\end{enumerate}

\end{enumerate}
\end{dn}

Tokens $i$ are considered as consistency witnesses for the consistency of the sets in $\con_{i}$. 
All requirements are very natural. Note that from Condition~\ref{dn-infofr}\eqref{dn-infofr-4}, that is, soundness, it particularly follows that for $i, j \in A$ and $X \in \con_{j}$,
\begin{equation}\label{eq-entcon}
X \vDash_{j} i \Rightarrow i R j.
\end{equation}

Sometimes a stronger version of Cut is needed which reverses the Interpolation Axiom.

\begin{lem}[\cite{sp25}]\label{lem-strong6}
Let $\bA$ be a continuous information frame. Then the following rule holds, for all $a, i, j \in A$, $X \in \con_{i}$ and $Y \in \con_{j}$,
\[
X \vDash_{i} (\{j\} \cup Y)  \kand Y \vDash_{j} a \Rightarrow X \vDash_{i} a.
\]
\end{lem}

Note next that from the global interpolation property \ref{dn-infofr}\eqref{dn-infofr-11} we in particular obtain that every local entailment relation $\vDash_{i}$ is interpolative.
\begin{lem}[\cite{sp25}]\label{lem-locint} 
Let $\bA$ be a continuous information frame. Then the following two statements hold for all $i \in A$ and $X, Y \fsubset A$ with $X \in \con_{i}$ and $X \vDash_{i} Y$:
\begin{enumerate}
\item\label{lem-locint-1}
$(\exists Z \in \con_{i}) X \vDash_{i} Z  \kand Z \vDash_{i} Y$.

\item\label{lem-locint-2}
$(\exists e \in A) X \vDash_{i} e \kand Y \in \con_{e}$.

\end{enumerate}
\end{lem}

\begin{dn}\label{dn-struth}
Let $\bA$ be a continuous information frame. 
\begin{enumerate}

\item\label{dn-struth-1}
$\bA$ is said to be \emph{strong} if the following Condition~\eqref{cn-S} holds for all $i \in A$ and $X \fsubset A$:
\begin{equation}\label{cn-S}
X \in \con_{i}\mbox{} \kand X \neq \{i\} \Rightarrow \{ i \} \vDash_{i} X. \tag{S}
\end{equation}

\item\label{dn-struth-2}
A token $\bt \in A$ is called \emph{truth element} if the following Condition~\eqref{cn-T} holds for all $i \in A$:
\begin{equation}\label{cn-T}
\emptyset \vDash_{i} \bt. \tag{T}
\end{equation}

\end{enumerate}
\end{dn}

As shown in \cite{sp25}, the domains generated from continuous information frames with truth element are pointed, and conversely, the canonical continuous information frames defined by  pointed domains have a truth element. In addition, they are strong.

Since information frames are families of local information systems the corresponding morphisms must be defined accordingly, i.e., as  families of locally operating approximable mappings.

 \begin{dn}\label{dn-am}
 Let  $\bA$ and $\bA'$ be continuous information frames. 
 \begin{enumerate}
 \item
 An \emph{approximable family} $\HH$ between $\bA$ and $\bA'$,  written $\HH = (H_{i})_{i \in A}:\bA \varpropto \bA'$, is a family of relations $H_{i} \subseteq \con_{i} \times A'$ satisfying for all $X, X' \in \con_{i}$, $Y \in \bigcup_{p \in A'} \con'_{p}$, $b, k \in A'$, and  $F \fsubset A'$ the following conditions, where $XH_{i}Y$ means that $XH_{i}c$, for all $c \in Y$:
 \begin{enumerate}
 \item \label{dn-am-1} $X H_{i} (\{k\} \cup Y) \kand Y \vDash'_{k} b \Rightarrow X H_{i} b$,
 
 \item\label{dn-am-2} $X' \supseteq X \kand X H_{i} b \Rightarrow X' H_{i} b$,
 
 \item\label{dn-am-3} $X \vDash_{i} X' \kand X' H_{i} b \Rightarrow X H_{i} b$,
 
 \item \label{dn-am-4}$i R j \kand X H_{i} b \Rightarrow X H_{j} b$,
 
 \item\label{dn-am-5} $X H_{i} F \Rightarrow  (\exists c \in A) (\exists e \in  A') (\exists U \in \con_{c}) (\exists V \in \con'_{e})\\
 \mbox{}\hfil\hfill  [X \vDash_{i} (\{c\} \cup U) \kand U H_{c} (\{e \} \cup V) \kand V \vDash'_{e} F]$.
 
 \end{enumerate}
 
 \item Let $\bA$ and $\bA'$, respectively, have truth elements $\bt$ and $\bt'$. Then $\appmapf{\HH}{\bA}{\bA'}$ \emph{respects truth elements}, if 
 
 \begin{enumerate}
 
 \item\label{dn-am-6}  $\emptyset H_{\bt} \bt'$.
 
 \end{enumerate}
 \end{enumerate}
 \end{dn}
 
 For $a \in A$, let $\bA_{a} = (A_{a}, \con_{a}, \vDash_{a})$ with $A_{a} = \bigcup \con_{a}$. Then $\bA_{a}$ is a continuous information system in the sense of Hoofman~\cite{ho93}. Moreover, for $a, b \in A$ with $a R b$, $S_{a}^{b} = \set{(X, c) \in \con_{a} \times A_{b}}{X \vDash_{b}c}$ is a continuous approximating mapping from $\bA_{a}$ to $\bA_{b}$ in the sense of  Hoofman.
 
Note that there is a corresponding version of Lemma~\ref{lem-amintan} for continuous information frames and approximating families.

\begin{lem}[\cite{sp21}]\label{pn-amint}
Let $\bA$ and $\bA'$ be continuous information frames. Then, for any family $(H_{i})_{i \in A}$ with $H_{i} \subseteq \con_{i} \times A'$, $X \in \con_{i}$, and $F \fsubset A'$, Condition~\ref{dn-am}\eqref{dn-am-5} is equivalent to the following Conditions~\eqref{pn-amint-1} and \eqref{pn-amint-2}:
\begin{enumerate}
\item\label{pn-amint-1}
$X H_{i} F \Rightarrow (\exists c \in A)(\exists U \in \con_{c})\, X \vDash_{i} (\{c\} \cup U) \kand U H_{c} F$,

\item\label{pn-amint-2}
$X H_{i} F \Rightarrow (\exists e \in A')(\exists V \in \con'_{e})\, X H_{i} (\{e\} \cup V) \kand V \vDash'_{e} F$.
\end{enumerate}
\end{lem}

Moreover, a strengthening of Condition~\ref{dn-am}\eqref{dn-am-3} can be derived which reverses the implication in the first statement of the preceding lemma.
\begin{lem}[\cite{sp21}]\label{lem-amstrong3}
Let $\HH$ be an approximable family between continuous information frames  $\bA$ and $\bA'$. Then for all $i, j \in A$, $X \in \con_{i}$, $Y \in \con_{j}$ and $b \in A'$,
\[
 X \vDash_{i} (\{j\} \cup Y) \kand Y H_{j} b \Rightarrow X H_{i} b.
\]
\end{lem}

Similar to the case of simplified continuous information systems, the family $(\vDash_{i})_{i \in A}$ of entailment relations in a continuous information frame $\bA$ is an approximable family, the idenity $\Id_{\bA}$ on $\bA$.

For $\nu = 1,2,3$, let $\bA^{(\nu)}$ be continuous information frames, $\appmapf{\GG}{\bA^{(1)}}{\bA^{(2)}}$ and $\appmapf{\HH}{\bA^{(2)}}{\bA^{(3)}}$. Define $\GG \circ \HH = ((\GG \circ \HH)_{i})_{i \in A^{(1)}}$ with
\begin{equation*}
X (\GG \circ \HH)_{i} a \Leftrightarrow (\exists e \in A^{(2)})(\exists V \in \con^{(2)}_{e})\, X G_{i} (\{e\} \cup V) \kand  V H_{e} a,
\end{equation*}
for $i \in A^{(1)}$, $X \in \con^{(1)}_{i}$ and $a \in A^{(3)}$. 

\begin{lem}\label{lem-amprop}
For $\nu = 1, 2, 3$, let $\bA^{(\nu)}$ be continuous information frames, $\GG: \bA^{(1)} \varpropto \bA^{(2)}$ and $\appmapf{\HH}{\bA^{(2)}}{\bA^{(3)}}$. Then the following statements hold:
\begin{enumerate}


\item\label{lem-amprop-2}
$\appmapf{\GG \circ \HH}{\bA^{(1)}}{\bA^{(3)}}$. 

\item\label{lem-amprop-3}
If $\GG$ and $\HH$ respect existing truth elements, the same does $\GG \circ \HH$.

\item\label{lem-amprop-4}
$(\vDash^{(1)}_{i})_{i \in A^{(1)}} \circ \GG = \GG \circ (\vDash^{(2)}_{j})_{j \in A^{(2)}} = \GG$.

\end{enumerate}
\end{lem}

Let $\CIF$ be the category of continuous information frames and approximable families,  and $\SIF$ be the full subcategory of strong continuous information frames.

\section{Information systems vs. information frames: a relationship}\label{sec-sysvsfr}

The goal of this and the following section is to investigate the relationship between simplified continuous information systems and  continuous information frames. 

Let $\bS = (S, \Con, \Vdash)$ be a simplified continuous information system and
\[
\FFF(\bS) = (A, (\hat{\con}_{i})_{i \in A}, (\mmodels_{i})_{i \in A})
\]
with
\begin{align*}
&A = \Con, \\
&\hat{\con}_{X} = \set{\XF \fsubset A}{\XF = \{X\} \kor (\forall Y \in \XF)\, X \Vdash Y}, \\
&\XF \mmodels_{X} Y \Leftrightarrow (\exists E \in \XF \cup \{X\})\, E \Vdash Y.
\end{align*}

\begin{thm}\label{thm-sisif}
Let $\bS$ be a simplified continuous information system. Then $\FFF(\bS)$ is a strong continuous information frame.
\end{thm}
For a proof of this theorem we have to verify the conditions in Definition~\ref{dn-infofr} and \ref{dn-struth}. Conditions~\ref{dn-infofr}\eqref{dn-infofr-1}, \ref{dn-infofr}\eqref{dn-infofr-2} and \eqref{cn-S} are immediate consequences of the definition of $\hat{\con}_{X}$ with $X \in \Con$. The remaining conditions are considered in the following lemmas.

\begin{lem}\label{lem-infofr-4}
Let $\XF \in \hat{\con}_{X}$ and $\YF \fsubset A$ with $\XF \mmodels_{X} \YF$. Then $\YF \in \hat{\con}_{X}$.
\end{lem}
\begin{pf}
We have that $\XF \in \hat{\con}_{X}$ and for all $Y \in \YF$ there exists $E \in \XF \cup \{X\}$ with $E \Vdash Y$. 
\begin{case}
$\XF \neq \{X\} \kand E \in \XF$.
\end{case}
Then $X \Vdash E$, by the definition of $\hat{\con}_{X}$. With \ref{dn-infosys}\eqref{dn-infosys-4} it follows that $X \Vdash Y$. Hence, $\YF \in \hat{\con}_{X}$.

\begin{case}
$E = X$.
\end{case}
Then $X \Vdash Y$. Again, we have that $\YF \in \hat{\con}_{X}$.
\end{pf}

\begin{lem}\label{lem-infofr-5}
Let $\XF, \YF \in \hat{\con}_{X}$ with $\YF \supseteq \XF$ and $\XF \mmodels_{X} Z$. Then $\YF \mmodels_{X} Z$.
\end{lem}
\begin{pf}
Let $\XF, \YF \in \hat{\con}_{X}$ with $\YF \supseteq \XF$ and $\XF \mmodels_{X} Z$. Then there is some $E \in \XF \cup \{X\}$ with $E \Vdash Z$. Since $\XF \subseteq \YF$, $E \in \YF  \cup \{X \}$ also holds. Hence, $\YF \mmodels_{X} Z$.
\end{pf}

\begin{lem}\label{lem-infofr-6}
Let $\XF \in \hat{\con}_{X}$, $\XF \mmodels_{X} \YF$ and $\YF \mmodels_{X} Z$. Then $\XF \mmodels_{X} Z$.
\end{lem}
\begin{pf}
Since $\YF \mmodels_{X} Z$, there is some $E \in \YF \cup \{X\}$ with $E \Vdash Z$. Similarly, there is some $C \in \XF \cup \{X\}$ with $C \Vdash E$. By \ref{dn-infosys}\eqref{dn-infosys-5} it follows that $C \Vdash Z$, which implies that $\XF \mmodels_{X} Z$.
\end{pf}

\begin{lem}\label{lem-infofr-7}
Let $\{X\} \in \hat{\con}_{Y}$. Then $\hat{\con}_{X} \subseteq \hat{\con}_{Y}$
\end{lem}
\begin{pf}
Without restriction let $X \neq Y$. Then, if $\{X\} \in \hat{\con}_{Y}$, we have that $Y \Vdash X$. Let $\ZF \in \hat{\con}_{X}$ with $\ZF \neq \{X\}$. It follows that for all $Z \in \ZF$,  $X \Vdash Z$. With \ref{dn-infosys}\eqref{dn-infosys-4} we obtain that also $Y \Vdash Z$, which shows that $\ZF \in \hat{\con}_{Y}$. 
\end{pf}

\begin{lem}\label{lem-infofr-9}
Let  $\{X\} \in \hat{\con}_{Y}$ and $\XF \in \hat{\con}_{X}$ with $\XF \mmodels_{X} Z$. Then $\XF \mmodels_{Y} Z$.
\end{lem}
\begin{pf}
Again, assume that $X \neq Y$. Since $\XF \mmodels_{X} Z$, there is some $E \in \XF \cup \{X\}$ with $E \Vdash Z$. Suppose that $E \neq X$. As $\XF \in \hat{\con}_{X}$, we then obtain that  $X \Vdash E$. It follows that in any case, $X \Vdash Z$. Since $\{X\} \in \hat{\con}_{Y}$, we have that $Y \Vdash X$. So, $Y \Vdash Z$. Consequently, there is some $F \in \XF \cup \{Y\}$ with $F \Vdash Z$, that is, $\XF \mmodels_{Y} Z$.
\end{pf} 

\begin{lem}\label{lem-infofr-11}
Let $\XF \mmodels_{X} \YF$. Then there are $Z \in \Con$ and $\ZF \in \hat{\con}_{Z}$ with $\XF \mmodels_{X} (\{Z\} \cup \ZF)$ and $\ZF \mmodels_{Z} \YF$.
\end{lem}
\begin{pf}
Assume that $\XF \mmodels_{X} \YF$ and let $Y \in \YF$. Then there is some $E \in \XF \cup \{X\}$ with $E \Vdash Y$. Let $c \in Y$. Because of \ref{dn-infosys}\eqref{dn-infosys-5} there is some $Z^{Y}_{c} \in \Con$ such that $E \Vdash Z^{Y}_{c}$ and $Z^{Y}_{c} \Vdash c$.  It follows that  $E \Vdash \bigcup\set{Z^{Y}_{c}}{c \in Y \kand Y \in \YF}$. Note that this union is finite, as $\YF$ and $Y$ are finite. Therefore, by \ref{dn-infosys}\eqref{dn-infosys-6}, there is some $Z \in \Con$ with $E \Vdash Z$ and $Z \supseteq \bigcup\set{Z^{Y}_{c}}{c \in Y \kand Y \in \YF}$. With \ref{dn-infosys}\eqref{dn-infosys-3} we now obtain that $Z \Vdash c$, for all $c \in Y$, that is $Z \Vdash Y$, for all $Y \in \YF$. Set $\ZF = \{Z\}$. Then $\ZF \in \hat{\con}_{Z}$. Moreover, $\XF \mmodels_{X} (\{Z\} \cup \ZF)$ and $\ZF \mmodels_{Z} \YF$.
\end{pf}

Next, we will study how the above construction can be extended to approximating mappings.  Let $\bS$ and $\bS'$ be simplified continuous information systems and $\appmap{H}{\bS}{\bS'}$.
Then, for $X \in \Con$, define $\hat{H}_{X} \subseteq \hat{\con}_{X} \times \Con'$ by
\[
\XF \hat{H}_{X} Y \Leftrightarrow (\exists E \in \XF \cup \{X\})\, E H Y
\]
and set $\FFF(H) = (\hat{H}_{X})_{X \in \Con}$.

\begin{lem}\label{lem-apmapf}
$\appmapf{\FFF(H)}{\FFF(\bS)}{\FFF(\bS')}$.
\end{lem}
We need to show that the assumptions in Definition~\ref{dn-am} are satisfied. This is done in the following lemmas.

\begin{lem}\label{lem-am-1}
Let $X \in \Con$, $\XF \in \hat{\con}_{X}$, $K \in \Con'$, $\YF \in \hat{\con'}_{K}$ and $Z \in \Con'$. If $\XF \hat{H}_{X} (\{K\} \cup \YF)$ and $\YF \mmodels'_{K} Z$, then $\XF \hat{H}_{X} Z$.
\end{lem}
\begin{pf}
Since $\YF \mmodels'_{K} Z$, there is some $Y_{Z} \in \YF \cup \{K\}$ with $Y_{Z} \Vdash' Z$; and because $\XF \hat{H}_{X} (\{K\} \cup \YF)$, we have that for all $Y \in \YF \cup \{K\}$ there is some $E_{Y} \in \XF \cup \{X\}$ with $E_{Y} H Y$. Then $E_{Y_{Z}} H Y_{Z}$ and $Y_{Z} \Vdash' Z$. With \ref{dn-appm}\eqref{dn-appm-1} it therefore follows that $E_{Y_{Z}} H Z$. Thus, $\XF \hat{H}_{X} Z$.
\end{pf}

\begin{lem}\label{lem-am-2}
Let $X \in \Con$,  $\XF, \tilde{\XF} \in \hat{\con}_{X}$  and $Y \in \Con'$ with $\tilde{\XF} \supseteq \XF$ and $\XF \hat{H}_{X} Y$. Then $\tilde{\XF} \hat{H}_{X} Y$.
\end{lem}
\begin{pf}
Since $\XF \hat{H}_{X} Y$, there is some $E \in \XF \cup \{X\}$ with $E H Y$. Then $E \in \tilde{\XF} \cup \{X\}$ also holds, as $\XF \subseteq \tilde{\XF}$. It follows that $\tilde{\XF} \hat{H}_{X} Y$.
\end{pf}

\begin{lem}\label{lem-am-3}
Let $X \in \Con$,  $\XF, \tilde{\XF} \in \hat{\con}_{X}$  and $Y \in \Con'$ with $\XF \mmodels_{X} \tilde{\XF}$ and $\tilde{\XF} \hat{H}_{X} Y$. Then $\XF \hat{H}_{X} Y$.
\end{lem}
\begin{pf}
It follows from the assumptions that there are $E \in \tilde{\XF} \cup \{X\}$ with $E H Y$ and  $Z \in \XF \cup \{X\}$ with $Z \Vdash E$. Thus, we have that $Z H Y$. Since $Z \in \XF \cup \{X\}$, this shows that $\XF \hat{H}_{X} Y$.  
\end{pf}

\begin{lem}\label{lem-am-4}
Let $X, Y \in \Con$, $\XF \in \hat{\con}_{X}$ and $Z \in \Con'$ with $\{X\} \in \hat{\con}_{Y}$ and $\XF \hat{H}_{X} Z$. Then $\XF \hat{H}_{Y} Z$.
\end{lem}
\begin{pf}
Without restrictions let $X \neq Y$. By the assumptions there is some $E \in \XF \cup \{X\}$ with $E H Z$. Moreover, $\hat{\con}_{X} \subseteq \hat{\con}_{Y}$ and hence $\XF \in \hat{\con}_{Y}$. If $E \in \XF$, it therefore follows that $\XF \hat{H}_{Y} Z$. If, on the other hand, $E = X$, then note that $\{ E \}\in \hat{\con}_{Y}$. Thus, $Y \Vdash E$. So, we have that $Y H Z$, that is, $\XF \hat{H}_{Y} Z$.
\end{pf}

\begin{lem}\label{lem-am-5}
Let $X \in \Con$,  $\XF \in \hat{\con}_{X}$ and $\FF \fsubset \Con'$ with $\XF \hat{H}_{X} \FF$. Then there are $K \in \Con$, $K' \in \Con'$, $\UF \in \hat{\con}_{K}$ and $\VF \in \hat{\con'}_{K'}$ so that $\XF \mmodels_{X} (\{K\} \cup \UF)$, $\UF \hat{H}_{K} (\{K'\} \cup \VF)$ and $\VF \mmodels'_{K'} \FF$.
\end{lem}
 \begin{pf}
With the assumption it follows that there is some $E_{Y} \in \XF \cup \{X\}$ with $E_{Y} H Y$, for every $Y \in \FF$. Hence, by Lemma~\ref{lem-amintan}, there are $Z_{Y} \in \Con$ and $Z'_{Y} \in \Con'$ with $E_{Y} \Vdash Z_{Y}$, $Z_{Y} H Z'_{Y}$ and $Z'_{Y} \Vdash' Y$. 
If $E_{Y} = X$, it follows that $X \Vdash Z_{Y}$. On the other hand, if $E_{Y} \neq X$, then also $\XF \neq \{X\}$. With Condition~\eqref{cn-S} it follows that $X \Vdash E_{Y}$, from which we obtain with Cut that $X \Vdash Z_{Y}$. So, in both cases we have that $X \Vdash \bigcup\set{Z_{Y}}{Y \in \FF}$. Because of Condition~\ref{dn-infosys}\eqref{dn-infosys-6} there is thus  some $\tilde{Z} \in \Con$ with $\tilde{Z} \supseteq  \bigcup\set{Z_{Y}}{Y \in \FF}$ and $X \Vdash \tilde{Z}$. As we have already seen $Z_{Y} H Z'_{Y}$. With \ref{dn-appm}\eqref{dn-appm-2} it therefore follows that $\tilde{Z} H \bigcup\set{Z'_{Y}}{Y \in \FF}$. By Condition~\ref{dn-appm}\eqref{dn-appm-5} there is some $\tilde{Z'} \in \Con'$ with $\tilde{Z'} \supseteq \bigcup\set{Z'_{Y}}{Y \in \FF}$ and $\tilde{Z} H \tilde{Z'}$. As seen above, $Z'_{Y} \Vdash' Y$. Therefore, $\tilde{Z'} \Vdash' \FF$, because of Weakening. Set $\ZF = \{\tilde{Z}\}$ and $\ZF' = \{\tilde{Z'}\}$. Then $\ZF \in \hat{\con}_{\tilde{Z}}$, $\ZF' \in \hat{\con'}_{\tilde{Z'}}$, $\XF \mmodels_{X} (\{\tilde{Z}\} \cup \ZF)$, $\ZF \hat{H}_{\tilde{Z}} (\{\tilde{Z'}\} \cup \ZF')$ and $\ZF' \mmodels'_{\tilde{Z'}} \FF$. 
 \end{pf}

\begin{pn}\label{pn-Ffunc}
$\fun{\FFF}{\SCIS}{\SIF}$ is a functor.
\end{pn}

In the next step we deal with the reverse situation. As we will see, every strong continuous information frame determines a continuous information system. The idea is to add the consistency witness $a \in A$ to the sets in $\con_{a}$, thus hiding it away.

Let $\bA =  (A, (\con_i)_{i \in A}, (\vDash_i)_{i \in A})$ be a strong continuous information frame and 
\[
\SSS(\bA) = (A, \Con, \Vdash)\]
 with
\begin{gather*}
\Con = \set{X \fsubset A}{(\exists a \in A) (\exists \ov{X} \in \con_{a})\, X = \ov{X} \cup \{a\}}, \\
X \Vdash c \Leftrightarrow  \ov{X} \vDash_{a} c\, \kor \{a\} \vDash_{a} c,
\end{gather*}
for $X \in \Con$ and $a \in A$, $\ov{X} \in \con_{a}$ with $X = \ov{X} \cup \{a\}$.

\begin{lem}\label{lem-infosys-6-1}
Let $X \in \Con$ and $F \fsubset A$ with $X \Vdash F$. Then there are $a \in A$ and $\ov{X} \in \con_{a}$ so that $X = \ov{X} \cup \{a\}$ and $\{a\}  \vDash_{a} F$.
\end{lem}
\begin{pf}
Since $X \Vdash F$, it follows that for every $c \in F$ there exist $a_{c} \in A$ and $X_{c} \in \con_{a_{c}}$ with $X = X_{c} \cup \{a_{c}\}$, and $X_{c} \vDash_{a_{c}}  c$ or $\{a_{c}\} \vDash_{a_{c}} c$. Due to Condition~\eqref{cn-S}, in both cases $\{a_{c}\} \vDash_{a_{c}} c$ holds.
Let $\ov{c} \in F$, $\ov{a} = a_{\ov{c}}$ and $\ov{X} = X_{\ov{c}}$. Then it follows for all $c \in F$ with $a_{c} = \ov{a}$ that  $\{\ov{a}\} \vDash_{\ov{a}} c$. Moreover, we have for all $c \in F$ with $a_{c} \neq \ov{a}$ that $a_{c} \in \ov{X}$. Since $\ov{X} \in \con_{\ov{a}}$, we obtain  for each of these $a_{c}$ that $\{a_{c}\} \in \con_{\ov{a}}$, that is, $a_{c} R \ov{a}$. Hence,  $\{a_{c}\} \vDash_{\ov{a}} c$. Furthermore, according to Condition~\eqref{cn-S}, $\{\ov{a}\} \vDash_{\ov{a}} a_{c}$. With Cut it now follows that $\{\ov{a}\} \vDash_{\ov{a}} c$. This shows that $\{\ov{a}\} \vDash_{\ov{a}} F$.
\end{pf}

\begin{thm}\label{thm-ifsis}
Let $\bA$ be a strong continuous information frame. Then $\SSS(\bA)$ is a continuous information system. 
\end{thm}
For the proof we have to verify Conditions~\ref{dn-infosys}\eqref{dn-infosys-1}-\eqref{dn-infosys-6}. This is done in a series of lemmas.

\begin{lem}\label{lem-infosys-1}
For all $a \in A$, $\{a\} \in \Con$.
\end{lem}
\begin{pf}
The statement is an immediate consequence of \ref{dn-infofr}\eqref{dn-infofr-1}.
\end{pf}

\begin{lem}\label{lem-infosys-2}
Let  $a \in A$ and $X \in \Con$ with $X \Vdash a$. Then $X \cup \{a\} \in \Con$. 
\end{lem}
\begin{pf}
Let $X \in \Con$. Then there are $i \in A$ and $\ov{X} \in \con_{i}$ with $X = \ov{X} \cup \{i\}$. 
Assume that $X \Vdash a$. Then $\ov{X} \vDash_{i} a$ or $\{i\} \vDash_{i} a$. Hence, $\{a\} \in \con_{i}$ by Condition~\ref{dn-infofr}\eqref{dn-infofr-4}. If $\ov{X} = \{i\}$, we thus have that $\{a\} \cup \{i\} \in \Con$, which means that $X \cup \{a\} \in \Con$. On the other hand, if $\ov{X} \neq \{i\}$, then $\{i\} \vDash_{i} \ov{X}$, as $\bA$ is strong.  In case that $\ov{X} \vDash_{i} a$ we therefore obtain with Cut  that also $\{i\} \vDash_{i} a$. Hence , we have in both cases that $\{i\} \vDash_{i} (\ov{X} \cup \{a\})$. By Soundness it follows that $\ov{X} \cup \{a\} \in \con_{i}$, which implies that $X \cup \{a\} \in \Con$.
\end{pf}

\begin{lem}\label{lem-infosys-3} 
Let $X, Y \in \Con$ with $X \subseteq Y$. If $X \Vdash a$, then also $Y \Vdash a$.
\end{lem}
\begin{pf}
Let $X, Y \in \Con$ with $X \subseteq Y$ and $X \Vdash a$. Then there are $i, j \in A$, $\ov{X} \in \con_{i}$ and $\ov{Y} \in \con_{j}$ so that $X = \ov{X} \cup \{i\}$ and $Y = \ov{Y} \cup \{j\}$. Moreover, $\ov{X} \vDash_{i} a$ or $\{i\} \vDash_{i} a$. Due to Condition~\eqref{cn-S}  it follows in both cases that $\{i\} \vDash_{i} a$. Since $X \subseteq Y$, we have that $i \in Y$.
\begin{case}
$i = j$.
\end{case}  
Then also $\{j\} \vDash_{j} a$ and therefore $Y \Vdash a$.

\begin{case}
$i \neq j$.
\end{case}
Then $i \in \ov{Y}$. By Condition~\eqref{cn-S}, $\{j\} \vDash_{j} \ov{Y}$. It follows that $\{j\} \vDash_{j} i$. With Lemma~\ref{lem-strong6} we hence have that  $\{j\} \vDash_{j} a$, that is, $Y \Vdash a$.
\end{pf}

\begin{lem}\label{lem-infosys-4} 
Let $a \in A$ and  $X, Y \in \Con$. If $X \Vdash Y$ and $Y \Vdash a$, then $X \Vdash a$.
\end{lem}
\begin{pf}
Let $a \in A$ and  $X, Y \in \Con$ so that  $X \Vdash Y$ and $Y \Vdash a$. By the definition of $\Con$ and Lemma~\ref{lem-infosys-6-1}, respectively, there are $i, j \in A$, $\ov{X} \in \con_{i}$ and $\ov{Y} \in \con_{j}$ such that $X = \ov{X} \cup \{i\}$,  $Y = \ov{Y} \cup \{j\}$, $\{i\} \vDash_{i} Y$ and $\{j\} \vDash_{j} a$, from which it follows with  Lemma~\ref{lem-strong6} that  $\{i\} \vDash_{i} a$, that is, $X \Vdash a$.
\end{pf}

\begin{lem}\label{lem-infosys-5}
Let $a \in A$ and $X \in \Con$ with $X \Vdash a$. Then there exists $Z \in \Con$ so that $X \Vdash Z$ and $Z \Vdash a$.
\end{lem}
\begin{pf}
Assume that $X \in \Con$ with $X \Vdash a$. Then there are $i \in A$ and $\ov{X} \in \con_{i}$ with $X = \ov{X} \cup \{i\}$, and $\ov{X} \vDash_{i} a$ or $\{i\} \vDash_{i} a$.  By \ref{dn-infofr}\eqref{dn-infofr-11} there are $e \in A$ and $U \in \con_{e}$ so that either $\ov{X} \vDash_{i} (\{e\} \cup U)$ and $U \vDash_{e} a$, or $\{i\} \vDash_{i} (\{e\} \cup U)$ and $U \vDash_{e} a$. Set $Z = U \cup \{e\}$. Then $Z \in \Con$, $X \Vdash Z$ and $Z \Vdash a$.
\end{pf}

\begin{lem}\label{lem-infosys-6-2}
Let $X \in \Con$ and $F \fsubset A$ with $X \Vdash F$. Then there is some $Z \in \Con$ with $Z \supseteq F$ and $X \Vdash Z$. 
\end{lem}
\begin{pf}
Let $X \in \Con$  and suppose that $X \Vdash F$. By Lemma~\ref{lem-infosys-6-1} there exist $i \in A$ and $\ov{X} \in \con_{i}$ with $X = \ov{X} \cup \{i\}$ and $\{i\} \vDash_{i} F$. With Condition~\ref{dn-infofr}\eqref{dn-infofr-11} it now follows that there are $e \in A$ and $U \in \con_{e}$ with $\{i\} \vDash_{i} (\{e\} \cup U)$ and  $U \vDash_{e} F$.  From the latter we obtain with Soundness that $F \in \con_{e}$. Hence, $\{e\} \cup F \in \Con$. Furthermore, we have that $\{i\} \vDash_{i} (F \cup \{e\})$. Set $Z = F \cup \{e\}$. Then $Z \supseteq F$, $Z \in \Con$ and $\{i\} \vDash_{i} Z$, that is, $X \Vdash Z$. 
\end{pf}

Next, we define how $\SSS$ operates on morphisms in $\SIF$.
Let to this end $\bA$, $\bA'$ be strong continuous information frames and $\appmapf{\HH}{\bA}{\bA'}$. For $X \in \Con$, let $a \in A$ and $\ov{X} \in \con_{a}$ with $X = \ov{X} \cup \{a\}$. Moreover, let $b \in A'$. Then define $\SSS(\HH) \subseteq \Con \times A'$ by
\[
X \SSS(\HH) b \Leftrightarrow \ov{X} H_{a} b\, \kor \{a\} H_{a} b.
\] 

\begin{lem}\label{lem-extmapset}
Let $X \in \Con$ and $F \fsubset A'$ with $X \SSS(\HH) F$. Then there exist $a \in A$ and $\ov{X} \in \con_{a}$ so that $X = \ov{X} \cup \{a\}$ and $\{a\} H_{a} F$.
\end{lem}
\begin{pf}
The proof proceeds as for Lemma~\ref{lem-infosys-6-1}.
\end{pf}

\begin{lem}\label{lem-apfapm}
$\appmap{\SSS(\HH)}{\SSS(\bA)}{\SSS(\bA')}$.
\end{lem}
\begin{pf}
We have to verify the conditions in Definition~\ref{dn-appm}.

\eqref{dn-appm-1}
Let $X \in \Con$, $Y \in \Con'$ and $b \in A'$ with $X \SSS(\HH) Y$ and $Y \Vdash' b$. Then there are $i \in A$, $j \in A'$, $\ov{X} \in \con_{i}$ and $\ov{Y} \in \con'_{j}$ so that $X = \ov{X} \cup \{i\}$ and $Y = \ov{Y} \cup \{j\}$. Moreover, $\ov{X} H_{i} (\ov{Y} \cup \{j\})$ or $\{i\} H_{i} (\ov{Y} \cup \{j\})$, and $\ov{Y} \vDash'_{j} b$ or $\{j\} \vDash'_{j} b$. We only consider the case that $\ov{X} H_{i} (\ov{Y} \cup \{j\})$ and $\ov{Y} \vDash'_{j} b$. In the other cases,  the procedure is completely analogously. By applying Condition~\ref{dn-am}\eqref{dn-am-1} we obtain in this case that $\ov{X} H_{i} b$, that is, $X \SSS(\HH) b$.

\eqref{dn-appm-2}
Let $X, Y \in \Con$ and $b \in A'$ with $X \subseteq Y$ and $X \SSS(\HH) b$. Then there are $i, j \in A$, $\ov{X} \in \con_{i}$ and $\ov{Y} \in \con_{j}$ so that $X = \ov{X} \cup \{i\}$ and $Y = \ov{Y} \cup \{j\}$. Moreover, $\ov{X} H_{i} b$ or $\{i\} H_{i} b$. With Conditions~\eqref{cn-S} and  \ref{dn-am}\eqref{dn-am-3} it follows in both cases that $\{i\} H_{i} b$. Since $X \subseteq Y$, we have that $i \in Y$.

\begin{case}
$i = j$.
\end{case}
Then also $\{j\} H_{j} b$ and therefore $Y \SSS(\HH) b$.

\begin{case}
$i \neq j$.
\end{case}
Then $i \in \ov{Y}$. By Condition~\eqref{cn-S}, $\{j\} \vDash_{j} \ov{Y}$. It follows that $\{j\} \vDash_{j} i$. With Lemma~\ref{lem-amstrong3} we therefore have that $\{j\} H_{j} b$, that is, $Y \SSS(\HH) b$.

\eqref{dn-appm-3}
Let $X, Y \in \Con$ and $b \in A'$ with $X \Vdash Y$ and $Y \SSS(\HH) b$. With Lemma~\ref{lem-infosys-6-1} it follows that there are $i \in A$ and $\ov{X} \in \con_{i}$ with $\{i\} \vDash_{i} Y$. Moreover,  there are $j \in A$ and $\ov{Y} \in \con_{j}$ with $Y = \ov{Y} \cup \{j\}$, and $\ov{Y} H_{j} b$ or $\{j\} H_{j} b$. In both cases it follows with Lemma~\ref{lem-amstrong3} that $\{i\} H_{i} b$, that is, $X \SSS(\HH) b$.

\eqref{dn-appm-4}
Let $X \in \Con$. Then there are $i \in A$ and $\ov{X} \in \con_{i}$ with $X = \ov{X} \cup \{i\}$. Assume that $X \SSS(\HH) b$, for $ b \in A'$. Then $\ov{X} H_{i} b$ or $\{i\} H_{i} b$. We only consider the first case. The second one can be handled analogously.

By Condition~\ref{dn-am}\eqref{dn-am-5} there are $c \in A$, $U \in \con_{c}$, $e \in A'$ and $V \in \con'_{e}$ so that $\ov{X} \vDash_{i} (\{c\} \cup U)$, $U H_{c} (\{e\} \cup V)$ and $V \vDash'_{e} b$. Set $Z = \{c\} \cup U$ and $Z' = \{e\} \cup V$. Then we have that $\ov{X} \vDash_{i} Z$ and $U H_{c} Z'$. In total, we therefore have that $X \Vdash Z$, $Z \SSS(\HH) Z'$ and $Z' \Vdash b$. 

\eqref{dn-appm-5}
Let $X \in \Con$ and $F \fsubset A'$ so that $X \SSS(\HH) F$. By Lemma~\ref{lem-extmapset} there exist  $i \in A$ and $\ov{X} \in \con_{i}$ with $X = \ov{X} \cup \{i\}$ and $\{i\} H_{i} F$. With Lemma~\ref{pn-amint}\eqref{pn-amint-2} it follows that there are $e \in A'$ and $V \in \con'_{e}$ so that $\{i\} H_{i} (\{e\} \cup V)$ and $V \vDash'_{e} F$. With Soundness we obtain that $F \in \con'_{e}$. Moreover, we have that $\{i\} H_{i} (\{e\} \cup F)$. Set $Z = F \cup \{e\}$. Then $Z \in \Con'$. Furthermore, $F \subseteq Z$ and $\{i\} H_{i} Z$. From the latter it follows that $X \SSS(\HH) Z$.
\end{pf}

\begin{pn}\label{pn-Sfunc}
$\fun{\SSS}{\SIF}{\CIS}$ is a functor.
\end{pn}

Let $\fun{\FFF'}{\CIS}{\SIF}$ be the restriction of the functor $\fun{\FFF}{\SCIS}{\SIF}$ to the full subcategory $\CIS$ of $\SCIS$.
In the remainder of this section we will show that the functors $\FFF'$ and $\SSS$ establish an equivalence between the categories $\SIF$ and $\CIS$.

For a category $\bC$ let $\ID_{\bC}$ be the identity functor on $\bC$. We first construct a natural isomorphism $\fun{\eta}{\ID_{\SIF}}{\FFF' \circ \SSS}$. Let to this end $\bA$ be a  strong continuous information frame. Then
\[
\FFF'(\SSS(\bA)) = (\Con, (\hat{\con}_{X})_{X \in \Con}, (\mmodels_{X})_{X \in \Con}),
\]
where for $X, Y \in \Con$ and $\XF \in \hat{\con}_{X}$,
\begin{align*}
\XF \mmodels_{X} &Y \\
\Leftrightarrow\mbox{} & (\exists E \in \XF \cup \{X\})\, E \Vdash Y \\
\Leftrightarrow\mbox{} & (\exists E \in \XF \cup \{X\}) (\forall c \in Y) (\exists e_{c} \in A)(\exists \ov{E_{c}} \in \con_{e})\, E = \ov{E_{c}} \cup \{e_{c}\} \kand \mbox{} \\
&(\ov{E_{c}} \vDash_{e_{c}} c\, \kor \{e_{c}\} \vDash_{e_{c}} c) \\
\Leftrightarrow\mbox{} & (\exists e \in A) (\exists \ov{E} \in \con_{e})\, \ov{E} \cup \{e\} \in \XF \cup \{X\} \kand \{e\} \vDash_{e} Y.
\end{align*}
Here, for the left-to-right implication of the last equivalence Lemma~\ref{lem-infosys-6-1} is applied.

For $i, a \in A$, $\ov{X}\in \con_{i}$, $K, Z \in \Con$ and $\XF \in \hat{\con}_{K}$ define
\begin{gather*}
\ov{X} P^{\bA}_{i} Z \Leftrightarrow \ov{X} \vDash_{i} Z, \\ 
\XF Q^{\bA}_{K} a \Leftrightarrow \XF \mmodels_{K} \{a\}
\end{gather*}
and set $P_{\bA} = (P^{\bA}_{i})_{i \in A}$ and $Q_{\bA} = (Q^{\bA}_{K})_{K \in \Con}$.
Then we have that
\begin{equation*}
\begin{split} 
\XF Q^{\bA}_{K} a &\Leftrightarrow (\exists E \in \XF \cup \{K\}) E \Vdash \{a\}\\  \label{eq-Q}
  &\Leftrightarrow (\exists e \in A)(\exists \ov{E} \in \con_{e})\, \ov{E} \cup \{e\} \in \XF \cup \{K\} \kand (\ov{E} \vDash_{e} a \kor \{e\} \vDash_{e} a).
 \end{split}
\end{equation*}

\begin{lem}\label{lem-propPQ}
\begin{enumerate}

\item \label{lem-propPQ-1}
$\appmapf{P_{\bA}}{\bA}{\FFF'(\SSS(\bA))}$.

\item \label{lem-propPQ-2}
$\appmapf{Q_{\bA}}{\FFF'(\SSS(\bA))}{\bA}$.

\item \label{lem-propPQ-3}
$P_{\bA} \circ Q_{\bA} = \Id_{\bA}$.

\item \label{lem-propPQ-4}
$Q_{\bA} \circ P_{\bA} = \Id_{\FFF'(\SSS(\bA))}$.

\end{enumerate}
\end{lem}
\begin{pf}
\eqref{lem-propPQ-1}
We have to show that the conditions in Definition~\ref{dn-am} hold.

\ref{dn-am}\eqref{dn-am-1} 
For $i \in A$, $\ov{X} \in \con_{i}$, $Y, Z \in \Con$ and $\ZF \in \hat{\con}_{Z}$ assume that $\ov{X} P^{\bA}_{i} (\{Z\} \cup \ZF)$ and $\ZF \mmodels_{Z} Y$. Then we have that $\ov{X} \vDash_{i} \bigcup (\ZF \cup \{Z\})$. Moreover, since  $\ZF \mmodels_{Z} Y$, there is some $K \in \ZF \cup \{Z\}$ with $K \Vdash Y$. With Lemma~\ref{lem-infosys-6-1} it follows that there exist $k \in A$ and $\ov{K} \in \con_{k}$ so that $K = \ov{K} \cup \{k\}$ and $\{k\} \vDash_{k} Y$. Hence $k  \in \bigcup (\ZF \cup \{Z\})$ and thus $\ov{X} \vDash_{i} k$. With Cut we therefore obtain that $\ov{X} \vDash_{i} Y$, that is, $\ov{X} P^{\bA}_{i} Y$.

The remaining conditions follow analogously by the corresponding properties of $(\vDash_{a})_{a \in A}$. 
For similar reasons also Statement~\eqref{lem-propPQ-2} holds.

\eqref{lem-propPQ-3}
Let $i, a \in A$ and $\ov{X} \in \con_{i}$. Then 
\begin{align*}
\ov{X} (P_{\bA} &\circ\mbox{} Q_{\bA})_{i} a  \\
\Leftrightarrow\mbox{} & (\exists Z \in \Con) (\exists \ZF \in \hat{\con}_{Z})\, \ov{X} P^{\bA}_{i} (\{Z\} \cup \ZF) \kand \ZF Q^{\bA}_{Z} a  \\
\Leftrightarrow\mbox{} &(\exists Z \in \Con) (\exists \ZF \in \hat{\con}_{Z})\, \ov{X} \vDash_{i} \bigcup (\ZF \cup \{Z\}) \kand \mbox{} \\
&(\exists k \in A)(\exists \ov{K} \in \con_{k})\, \ov{K} \cup \{k\} \in \ZF \cup \{Z\} \kand (\ov{K} \vDash_{k} a\, \kor \{k\} \vDash_{k} a) \\
\Leftrightarrow\mbox{} & \ov{X} \vDash_{i} a,
\end{align*}
where the left-to-right implication in the last equivalence follows with Lemma~\ref{lem-strong6}. For the reverse direction use Interpolation to obtain $k \in A$ and $\ov{K} \in \con_{k}$ with $\ov{X} \vDash_{i} (\{k\} \cup \ov{K})$ and $\ov{K} \vDash_{k} a$. Set $Z = \ov{K} \cup \{k\}$ and $\ZF = \{Z\}$. Then $Z \in \Con$ and $\ZF \in \hat{\con}_{Z}$.

\eqref{lem-propPQ-4}
Let $X, Y \in \Con$ and $\XF \in \hat{\con}_{X}$. Then 
\begin{align*}
\XF (Q_{\bA} \circ\mbox{} &P_{\bA})_{X} Y \\
\Leftrightarrow\mbox{} &(\exists a \in A) (\exists \ov{Z} \in \con_{a})\, \XF Q^{\bA}_{X} (\{a\} \cup \ov{Z}) \kand \ov{Z} P^{\bA}_{a} Y \\
 \Leftrightarrow\mbox{} & (\exists a \in A) (\exists \ov{Z} \in \con_{a}) (\exists e \in A) (\exists \ov{E} \in \con_{e})\, \ov{E} \cup \{e\} \in \XF \cup \{X\} \kand \mbox{} \\
 & (\ov{E} \vDash_{e} (\{a\} \cup \ov{Z}) \kor \{e\} \vDash_{e} (\{a\} \cup \ov{Z})) \kand \ov{Z} \vDash_{a} Y \\
\Leftrightarrow\mbox{} & (\exists e \in A) (\exists \ov{E} \in \con_{e})\, \ov{E} \cup \{e\} \in \XF \cup \{X\} \kand (\ov{E} \vDash_{e} Y \kor \{e\} \vDash_{e} Y) \\
\Leftrightarrow\mbox{} & \XF \mmodels_{X} Y.
\end{align*}
Here, the left-to-right implication of the next-to-last equivalence follows with Lemma~\ref{lem-strong6}. For the converse direction use Interpolation.
\end{pf}

Set $\eta_{\bA} = P_{\bA}$. We want to show that $\eta$ is a natural transformation.

\begin{lem}\label{lem-treta}
Let $\bA'$ be a further strong continuous information frame and $\appmapf{\HH}{\bA}{\bA'}$. Then the following statements hold:
\begin{enumerate}

\item\label{lem-treta-1}
Let $\tilde{\HH} =  \FFF'(\SSS(\HH))$. Then for $X \in \Con$, $\XF \in \hat{\con}_{X}$ and $Y' \in \Con'$, 
\[
\XF \tilde{H}_{X} Y \Leftrightarrow (\exists e \in A) (\exists \ov{E} \in \con_{e})\, \{e\} \cup \ov{E} \in \XF \cup \{X\} \kand  \{e\} H_{e} Y.
\]

\item\label{lem-treta-2}
$
 \HH \circ P_{\bA'} = P_{\bA} \circ \FFF'(\SSS(\HH)).
$
\end{enumerate}
\end{lem}
\begin{pf}
\eqref{lem-treta-1} follows easily with Lemma~\ref{lem-extmapset}.

For \eqref{lem-treta-2} let  $a \in A$, $\ov{X} \in \con_{a}$ and $Y' \in \Con'$. Then 
\begin{align*}
\ov{X} (\HH \circ P_{\bA'})_{a} Y' 
&\Leftrightarrow (\exists k' \in A')(\exists \ov{Z'} \in \con'_{k'})\, \ov{X} H_{a} (\{k'\} \cup \ov{Z'}) \kand \ov{Z'} P^{\bA'}_{k'} Y' \notag \\
&\Leftrightarrow (\exists k' \in A')(\exists \ov{Z'} \in \con'_{k'})\, \ov{X} H_{a} (\{k'\} \cup \ov{Z'}) \kand \ov{Z'} \vDash'_{k'} Y'.  \\
&\Leftrightarrow \ov{X} H^{\bA}_{a} Y',
\end{align*}
where the last equivalence follows with Lemma~\ref{lem-amstrong3} and Lemma~\ref{pn-amint}\eqref{pn-amint-2}, respectively.

On the other hand we have
\begin{align}
\ov{X} (P_{\bA} \circ \tilde{\HH})_{a} Y'
\Leftrightarrow\mbox{} &(\exists Z \in \Con)(\exists \ZF \in \hat{\con}_{Z})\, \ov{X} P^{\bA}_{a} (\{Z\} \cup \ZF) \kand \ZF \tilde{H}_{Z} Y' \notag\\
\Leftrightarrow\mbox{} &(\exists Z \in \Con)(\exists \ZF \in \hat{\con}_{Z})\, \ov{X} \vDash_{a} \bigcup (\{Z\} \cup \ZF) \kand\mbox{} \label{eq-st1}\\
      &(\exists e \in A) (\exists \ov{E} \in \con_{e})\, \{e\} \cup \ov{E} \in \{Z\} \cup\ZF \kand  \{e\} H_{e} Y'  \notag\\
 \Leftrightarrow\mbox{} &(\exists e \in A) (\exists \ov{E} \in \con_{e})\, \ov{X} \vDash_{a} (\{e\} \cup \ov{E}) \kand \{e\} H_{e} Y'  \label{eq-spec}\\
 \Leftrightarrow\mbox{} &\ov{X} H_{a} Y',\notag
\end{align}
where the last equivalence follows with Lemma~\ref{lem-amstrong3} and Lemma~\ref{pn-amint}\eqref{pn-amint-1}, respectively. The left-to-right implication of Equivalence~\eqref{eq-spec} is obvious; for the reverse direction choose $Z = \ov{E} \cup \{e\}$ and $\ZF = \{Z\}$. In Equivalence~\eqref{eq-st1} the first statement of this lemma is used.
\end{pf}

Let us now summarise what we have shown so far.

\begin{pn}\label{pn-ciftofs}
$\fun{\eta}{\ID_{\SIF}}{\FFF' \circ \SSS}$ is a natural isomorphism.
\end{pn}

Next, we show that there is also a natural isomorphism $\fun{\tau}{\ID_{\CIS}}{\SSS \circ \FFF'}$. Let to this end $\bS = (S, \Con, \Vdash)$ be a continuous information system. Then 
\[
\SSS(\FFF'(\bS)) = (\Con, \hat{\Con}, \Vvdash),
\]
where
\begin{gather*}
\hat{\Con} = \set{\XF \fsubset \Con}{(\exists X \in \Con) (\exists \ov{\XF} \in \hat{\Con}_{X})\, \XF = \ov{\XF} \cup \{X\}}, \\
\XF \Vvdash Y \Leftrightarrow (\exists E \in \XF)\, E \Vdash Y.
\end{gather*}

Define $S_{\bS} \subseteq \Con \times \Con$ and $T_{\bS} \subseteq \hat{\Con} \times S$ by
\begin{gather*}
X S_{\bS} Y \Leftrightarrow X \Vdash Y, \\
\XF T_{\bS} a \Leftrightarrow (\exists E \in \XF)\, E \Vdash a.
\end{gather*}

\begin{lem}\label{lem-profST}
\begin{enumerate}

\item\label{lem-profST-1}
$\appmap{S_{\bS}}{\bS}{\SSS(\FFF'(\bS))}$.

\item\label{lem-profST-2}
$\appmap{T_{\bS}}{\SSS(\FFF'(\bS))}{\bS}$.

\item\label{lem-profST-3}
$S_{\bS} \circ T_{\bS} = \Id_{\bS}$.

\item\label{lem-profST-4}
$T_{\bS} \circ S_{\bS} = \Id_{\SSS(\FFF'(\bS))}$.

\end{enumerate}
\end{lem}
\begin{pf}
\eqref{lem-profST-1}
We have to show that $S_{\bS}$ satisfies the conditions in Definition~\ref{dn-appm}. Conditions~\eqref{dn-appm-1}-\eqref{dn-appm-3} are obvious and Condition~\eqref{dn-appm-5} follows similarly to Condition~\eqref{dn-appm-4}. So, we only look at this one. 

Let $X, Y \in \Con$ so that $X S_{\bS} Y$. Then $X \Vdash Y$. Hence, for every $y \in Y$, there are $Z_{y}, Z'_{y} \in \Con$ with $X \Vdash Z_{y}$, $Z_{y} \Vdash Z'_{y}$ and $Z'_{y} \Vdash y$. It follows that $X \Vdash \bigcup\set{Z_{y}}{y \in Y}$. By Condition~\ref{dn-infosys}\eqref{dn-infosys-5} there is thus some $Z \in \Con$ with $Z \supseteq \bigcup\set{Z_{y}}{y \in Y}$  and $X \Vdash Z$. With Weakening it follows that $Z \Vdash \bigcup\set{Z_{y}}{y \in Y}$. By repeating this step we obtain some $Z' \in \Con$ such that $Z \Vdash Z'$ and also $Z' \Vdash Y$.

\eqref{lem-profST-2}
Now, we must show that $T_{\bS}$ satisfies the conditions in Definition~\ref{dn-appm}. Again, we only consider Condition~\eqref{dn-appm-4}. 

Let $\XF \in \hat{\Con}$ and $a \in S$ with $\XF T_{\bS} a$. Then there is some $E \in \XF$ so that $E \Vdash \{a\}$. As in the previous step it follows that there are $Z, Z' \in \Con$ such that $E \Vdash Z$, $Z \Vdash Z'$ and $Z' \Vdash \{a\}$. Set $\ZF = \{Z\}$. Then $\ZF \in \hat{\Con}$ and we have that  $\XF \Vvdash \ZF$, $\ZF T_{\bS} Z'$ and $Z' \Vdash \{a\}$.

\eqref{lem-profST-3}
Let $a \in S$ and $X \in \Con$. Then we have that 
\begin{align*}
X (S_{\bS} \circ T_{\bS}) a
& \Leftrightarrow (\exists \ZF \in \hat{\Con})\, X S_{\bS} \ZF \kand \ZF T_{\bS} a \\
& \Leftrightarrow (\exists \ZF \in \hat{\Con}) (\forall Z \in \ZF)\, X \Vdash Z \kand\, (\exists E \in \ZF)\, E \Vdash a \\
& \Leftrightarrow X \Vdash a.
\end{align*}
Here, the  left-to-right implication of the last equivalence is obvious. For the reverse implication set $\ZF = \{E\}$, where the existence of $E$ with $X \Vdash E$ and $E \Vdash a$ follows by Condition~\ref{dn-infosys}\eqref{dn-infosys-4}.

\eqref{lem-profST-4}
Let $\XF \in \hat{\Con}$ and $Y \in \Con$. Then 
\begin{align*}
\XF (T_{\bS} \circ S_{\bS}) Y 
&\Leftrightarrow (\exists Z \in \Con)\, \XF T_{\bS} Z \kand Z S_{\bS} Y \\
&\Leftrightarrow (\exists Z \in \Con) (\exists E \in \XF)\, E \Vdash Z \kand Z \Vdash Y \\
&\Leftrightarrow (\exists E \in \XF)\, E \Vdash Y \\
&\Leftrightarrow \XF \Vvdash Y. 
\end{align*}
\end{pf}

Set $\tau_{\bS} = S_{\bS}$. We want to show that $\tau$ is a natural transformation.
Let to this end $\bS'$ be a further continuous information system and $\appmap{H}{\bS}{\bS'}$. Set $\hat{H} = \SSS(\FFF'(H))$. Then we have for $\XF \in \hat{\Con}$ and $Y \in \Con'$ that
\[
\XF \hat{H} Y \Leftrightarrow (\exists E \in \XF)\, E H Y.
\]

\begin{lem}\label{lem-trtau}
$H \circ S_{\bS'}  = S_{\bS} \circ \hat{H}$.
\end{lem}
\begin{pf}
Let $X \in \Con$ and $Y \in \Con'$. Then 
\begin{align*}
X (H \circ S_{\bS'}) Y 
&\Leftrightarrow (\exists Z' \in \Con')\, X H Z' \kand Z' S_{\bS'} Y \\
&\Leftrightarrow  (\exists Z' \in \Con')\, X H Z' \kand Z' \Vdash' Y \\
&\Leftrightarrow X H Y,
\end{align*}
where the last equivalence follows with Condition~\ref{dn-appm}\eqref{dn-appm-1} and Lemma~\ref{lem-amintan}\eqref{lem-amintan-2}, respectively.

Moreover, we have 
\begin{align*}
X (S_{\bS} \circ \hat{H}) Y
&\Leftrightarrow (\exists \ZF \in \hat{\Con})\, X S_{\bS} \ZF \kand \ZF \hat{H} Y \\
&\Leftrightarrow (\exists \ZF \in \hat{\Con}) (\forall Z \in \ZF)\, X \Vdash Z \kand (\exists E \in \ZF)\, E H Y \\
&\Leftrightarrow (\exists E \in \Con)\, X \Vdash E \kand E H Y \\
&\Leftrightarrow X \Vdash Y.
\end{align*}
The existence of $E$ in the right-to-left implication of the last equivalence follows with Lemma~\ref{lem-amintan}\eqref{lem-amintan-1}. In the same direction of the next-to-last equivalence choose $\ZF = \{E\}$.
\end{pf}

Let us again summarise what we have achieved in this step.

\begin{pn}\label{pn-scistosf}
$\fun{\tau}{\ID_{\CIS}}{\SSS \circ \FFF'}$ is a natural isomorphism.
\end{pn}

Putting Propositions~\ref{pn-ciftofs} and \ref{pn-scistosf} together, we obtain what we were aiming for in this section.

\begin{thm}\label{thm-cifcis}
The categories $\CIS$ and $\SIF$ are equivalent.
\end{thm}

In the remainder of this section we want to show that the functor 
\[
\fun{\SSS \circ \FFF}{\SCIS}{\CIS}
\]
 is left-adjoint to the inclusion functor $\fun{\UUU}{\CIS}{\SCIS}$. Let to this end $\bS = (S, \Con, \Vdash)$ be a simplified continuous information system. Then we have as above that
\[
\SSS(\FFF(\bS)) = (\Con, \hat{\Con}, \Vvdash),
\]
where
\begin{gather*}
\hat{\Con} = \set{\XF \fsubset \Con}{(\exists X \in \Con) (\exists \ov{\XF} \in \hat{\Con}_{X})\, \XF = \ov{\XF} \cup \{X\}}, \\
\XF \Vvdash Y \Leftrightarrow (\exists E \in \XF)\, E \Vdash Y.
\end{gather*}

Define $S'_{\bS} \subseteq \Con \times \Con$ and $T'_{\bS} \subseteq \hat{\Con} \times S$ by
\begin{gather*}
X S'_{\bS} Y \Leftrightarrow X \Vdash Y, \\
\XF T'_{\bS} a \Leftrightarrow (\exists E \in \XF)\, E \Vdash a.
\end{gather*}

Since Condition~\ref{dn-infosys}\eqref{dn-infosys-2} has never been used in the proofs of lemmas~\ref{lem-profST} and \ref{lem-trtau}, it follows again that $\appmap{S'_{\bS}}{\bS}{\SSS(\FFF(\bS))}$, $\appmap{T'_{\bS}}{\SSS(\FFF(\bS))}{\bS}$ and both approximable mappings are inverse to each other. Moreover, $\fun{\tau'}{\ID_{\SCIS}}{\UUU \circ (\SSS \circ \FFF)}$ with $\tau'_{\bS} = S'_{\bS}$ is a natural isomorphism. Hence, $\SSS \circ \FFF$ is left-adjoint to $\UUU$.

\begin{thm}\label{thm-reflCIS}
The category $\CIS$ is a full reflective subcategory of  $\SCIS$.
\end{thm}

Let $\tau''_{\bS} = S'_{\bS}$, for continuous information systems $\bS$. Then we also have that $\fun{\tau''}{\ID_{\CIS}}{(\SSS \circ \FFF) \circ \UUU}$ is a natural isomorphism.

\begin{thm}\label{thm-isoCIS}
The categories $\CIS$ and $\SCIS$ are equivalent.
\end{thm}

\section{Strong continuous information frames}\label{sec-strong}

The category $\SIF$ of strong continuous information frames is a full subcategory of $\CIF$. As we will show in this section, each continuous information frame can be turned into a strong one.

Let $\bA = (A, (\con_{i})_{i \in A}, (\vDash_{i})_{i \in A})$ be a continuous information frame and define
\begin{align*}
&\tilde{A} = \bigcup\set{\{a\} \times \con_{a}}{a \in A}, \\
&\tilde{\con}_{(a, X)} = \{\{(a, X)\}\} \cup \set{\YF \fsubset \tilde{A}}{(\forall (b, Y) \in \YF)\, X \vDash_{a} \{b\} \cup Y},  \\
&\XF \tdash_{(a, X)} (e, V) \Leftrightarrow (\exists (c, Z) \in \XF \cup \{(a, X)\})\,   Z \vDash_{c} \{e\} \cup V. 
\end{align*}
Set 
\[
\TTT(\bA) = (\tilde{A}, (\tilde{\con}_{X})_{X \in \tilde{A}}, (\tdash_{X})_{X \in \tilde{A}}).
\]

\begin{thm}\label{thm-cs}
$\TTT(\bA)$ is a strong continuous information frame.
\end{thm}
\begin{pf}
We have to verify Condition~\eqref{cn-S} and the conditions in Definition~\ref{dn-infofr}. Let to this end $(a, X) \in \tilde{A}$.

\eqref{cn-S}
Assume that $\XF \in \tilde{\con}_{X}$ with $\XF \neq \{(a, X)_{}\}$. We have to show that for all $(c, Z) \in \XF$, $\{(a, X) \} \tdash_{(a, X)} (c, Z)$, that is, $X \vDash_{a} \{c\} \cup Z$, which holds by the Definition of $\tilde{\con}_{X}$.

\eqref{dn-infofr-1} also holds by the Definition on $\tilde{\con}_{X}$ and
\eqref{dn-infofr-2} is obvious.

\eqref{dn-infofr-4} 
Let $\XF \in \tilde{\con}_{(a, X)}$ and $\XF \tdash_{(a, X)} \YF$. Then, for each $(b, Y) \in \YF$, there is some $(c^{b}_{Y}, Z^{b}_{Y}) \in \XF \cup \{(a, X)\}$ with $Z^{b}_{Y} \vDash_{c^{b}_{Y}} (\{b\} \cup Y)$. Since $\XF \in \tilde{\con}_{(a, X)}$, we have of all $(c^{b}_{Y}, Z^{b}_{Y}) \in \XF$ that $X \vDash_{a} (\{c^{b}_{Y}\} \cup  Z^{b}_{Y})$; similarly, if $(c^{b}_{Y}, Z^{b}_{Y}) = (a, X)$. So, we obtain with Lemma~\ref{lem-strong6} that $X \vDash_{a} (\{b\} \cup Y)$, for all $Y \in  \YF$. Thus, $\YF \in \tilde{\con}_{(a, X)}$. 

\eqref{dn-infofr-5}
Let $(e, V) \in \tilde{A}$, $\XF, \YF \in \tilde{\con}_{(a, X)}$ with $\XF \subseteq \YF$, and assume that $\XF \tdash_{(a, X)} (e, V)$. Then there is some $(c, Z) \in \XF \cup \{(a, X)\}$ so that $Z \vDash_{c} (\{e\} \cup V)$. Then $(c, Z) \in \YF \cup \{(a, X)\}$ also holds. Hence, $\YF \tdash_{(a, X)} V$.

\eqref{dn-infofr-6}
Suppose that $\XF, \YF \in \tilde{\con}_{(a, X)}$ and $(e, V) \in \tilde{A}$ so that $\XF \tdash_{(a, X)} \YF$ and $\YF \tdash_{(a, X)} (e, V)$. Then there is some $(c, Z) \in \YF \cup \{(a, X)\}$ with $Z \vDash_{c} (\{e\} \cup V)$ and some $(d, U) \in \XF \cup \{(a, X)\}$ with $U \vDash_{d} (\{c\} \cup Z)$. With Lemma~\ref{lem-strong6} it follows that $U \vDash_{d} (\{e\} \cup V)$. Hence, $\XF \tdash_{(a, X)} (e, V)$.

\eqref{dn-infofr-7}
Let  $(b, Y) \in \tilde{A}$ so that $\{(a, X)\} \in \tilde{\con}_{(b, Y)}$. Without restriction assume that $(a, X) \neq (b, Y)$. Then $Y \vDash_{b} (\{a\} \cup X)$.  Now, let $\XF \in \tilde{\con}_{(a, X)}$. If $\XF = \{(a, X)\}$, then $\XF \in \tilde{\con}_{(b, Y)}$, by assumption. So, suppose that $\XF \neq \{(a, X)\}$ and let $(c, Z) \in \XF$. Then $X \vDash_{a} (\{c\} \cup Z)$. Since we also have that $Y \vDash_{b} (\{a\} \cup X)$, it follows that $Y \vDash_{b} (\{c\} \cup Z)$. In total we obtain that $\XF \in \tilde{\con}_{(b, Y)}$.

\eqref{dn-infofr-9}
Let $\{(a, X)\} \in \tilde{\con}_{(b, Y)}$. Without restriction let again $(a, X) \neq (b, Y)$.  Then $Y \vDash_{b} (\{a\} \cup X)$. Assume that $\XF \in \tilde{\con}_{(a, X)}$ and $(e, V) \in \tilde{A}$ with $\XF \tdash_{(a, X)} (e, V)$.  Then there is some $(c, Z) \in \XF \cup \{(a, X)\}$ so that $Z \vDash_{c} (\{e\} \cup V)$. Set $(c', Z') = (c, Z)$, if $(c, Z) \in \XF$, and $(c', Z') = (b, Y)$, if $(c, Z) = (a, X)$. Then we have that $(c', Z') \in \XF \cup \{(b, Y)\}$ and $Z' \vDash_{c'} (\{e\} \cup V)$. Hence, $\XF \tdash_{(b, Y)} (e, V)$.

\eqref{dn-infofr-11}
Let $\XF \in \tilde{\con}_{(a, X)}$ and $\VF \fsubset \tilde{A}$ with $\XF \tdash_{(a, X)} \VF$. Then, for each $(e, V) \in \VF$ there is some $(c^{e}_{V}, Z^{e}_{V}) \in \XF \cup \{(a, X)\}$ with $Z^{e}_{V} \vDash_{c^{e}_{V}} (\{e\} \cup V)$. If $(c^{e}_{V}, Z^{e}_{V})  \in \XF$, it follows that $X \vDash_{a} (\{c^{e}_{V}\} \cup Z^{e}_{V})$, as $\XF \in \tilde{\con}_{(a, X)}$. In total we thus have that $X \vDash_{a} \bigcup\set{\{e\} \cup V}{(e, V) \in \VF}$. Because of Interpolation there are $d \in A$ and $U \in \con_{d}$ with $X \vDash_{a} (\{d\} \cup U)$ and $U \vDash_{d} \bigcup\set{\{e\} \cup V}{(e, V) \in \VF}$. Then $\{(d, U)\} \in \tilde{\con}_{(d, U)}$ and we have $\{(d, U)\} \tdash_{(d, U)} \VF$. Since, moreover, $X \vDash_{a} (\{d\} \cup U)$, we further obtain that $\XF \tdash_{(a, X)} (d, U)$. So, we have that there is some $\UF \in \tilde{\con}_{(d, U)}$ with $\XF \tdash_{(a, X)} \UF$ and $\UF \tdash_{(d, U)} \VF$, namely $\UF = \{(d, U)\}$.
\end{pf}

In a further step, we will now extend the construction to approximable families. Let to this end $\bA'$ be another continuous information frame and $\appmapf{\HH}{\bA}{\bA'}$. For $(a, X) \in \tilde{A}$, $\XF \in \tilde{\con}_{(a, X)}$ and $(b, Y) \in \tilde{A}'$ define 
\[
\XF \tilde{H}_{(a, X)} (b, Y) \Leftrightarrow (\exists (c, Z) \in \XF \cup \{(a, X)\})\, Z H_{c} (\{b\} \cup Y).
\]

\begin{lem}\label{lem-csf}
$\appmapf{\tilde{\HH} = (\tilde{H}_{(a, X)})_{(a, X) \in \tilde{A}}}{\TTT(\bA)}{\TTT(\bA')}$.
\end{lem}
\begin{pf}
We need to verify the conditions in Definition~\ref{dn-am}. Let to this end $(a, X) \in \tilde{A}$ and $\XF \in \tilde{\con}_{(a, X)}$.

\eqref{dn-am-1}
Assume that $\XF \tilde{H}_{(a, X)} (\{(k, K)\} \cup \KF)$ and $\KF \tdash'_{(k, K)} (b, Y)$, where $(k, K)$, $(b, Y) \in \tilde{A}'$ and $\KF \in \tilde{\con}'_{(k, K)}$. 
Then there is some $(d, U) \in \KF \cup \{(k, K)\}$ with $U \vDash'_{d} (\{b\} \cup Y)$. Moreover, as $\XF \tilde{H}_{(a, X)} (\{(k, K)\} \cup \KF)$, we have that $\XF \tilde{H}_{(a, X)} (d, U)$. Thus, there is some $(c, Z) \in \XF \cup \{(a, X)\}$ so that $Z H_{c} (\{d\} \cup U)$. It follows that $Z H_{c} (\{b\} \cup Y)$, which shows that $\XF \tilde{H}_{(a, X)} (b, Y)$.

\eqref{dn-am-2}
is obvious.

\eqref{dn-am-3}
Let in addition $\XF' \in \tilde{\con}_{(a, X)}$ so that $\XF \tdash \XF'$ and $\XF' \tilde{H}_{(a, X)} (b, Y)$. Then there is some $(c, Z) \in \XF' \cup \{(a, X)\}$ with $Z H_{c} (\{b\} \cup Y)$. Since we also have that $\XF \tdash (c, Z)$, there is some $(d, U) \in \XF \cup \{(a, X)\}$ such that $U \vDash_{d} (\{c\} \cup Z)$. It follows that $\{c\} \in \con_{d}$. Hence, we have that $Z H_{d} (\{b\} \cup Y)$. With \ref{dn-am}\eqref{dn-am-4} we now obtain that $U H_{d} (\{b\} \cup Y)$, that is, $\XF \tilde{H}_{(a, X)} (b, Y)$. 

\eqref{dn-am-4}
Let $\{(a, X)\} \in \tilde{\con}_{(e, V)}$ and without restriction let $(a, X) \neq (e, V)$. Assume that $\XF \tilde{H}_{(a, X)} (b, Y)$. Then there is some $(d, U) \in \XF \cup \{(a, X)\}$ with $U H_{d} (\{b\} \cup Y)$. Since $\{(a, X)\} \in \tilde{\con}_{(e, V)}$, we have that $\XF \in \tilde{\con}_{(e, V)}$. So, in particular $\{(d, U)\} \in \tilde{\con}_{(e, V)}$. Thus, $V \vDash_{e} (\{d\} \cup U)$. By Lemma~\ref{lem-amstrong3}, we obtain that $V H_{e} (\{b\} \cup Y)$, which means that $\XF \tilde{H}_{(e, V)} (b, Y)$. 

 \eqref{dn-am-5}
 Let $\FF \fsubset \tilde{A}'$ and assume that $\XF \tilde{H}_{(a, X)} \FF$. Then, for every $(b, Y) \in \FF$ there is some $(c^{b}_{Y}, Z^{b}_{Y}) \in \XF \cup \{(a, X)\}$ with $Z^{b}_{Y} H_{c^{b}_{Y}} (\{b\} \cup Y)$. If $\XF = \{(a, X)\}$ then $(c^{b}_{Y}, Z^{b}_{Y}) = (a, X)$ and hence $X H_{a}  (\{b\} \cup Y)$. Otherwise, $X \vDash_{a} (\{c^{b}_{Y}\} \cup Z^{b}_{Y})$, from which we also obtain that $X H_{a}  (\{b\} \cup Y)$. So, in both cases we have that $X H_{a} \bigcup\set{Y \cup \{b\}}{(b, Y) \in \FF}$. With \ref{dn-am}\eqref{dn-am-5} it follows that there are $d \in A$, $e \in A'$, $U \in \con_{d}$ and $V \in \con'_{e}$ with $X \vDash_{a} (\{d\} \cup U)$, $U H_{d} (\{e\} \cup V)$ and $V \vDash'_{e} \bigcup\set{Y \cup \{b\}}{(b, Y) \in \FF}$. Set $\UF = \{(d, U)\}$ and $\VF = \{(e, V)\}$. Then $\UF \in \tilde{\con}_{(d, U)}$ and $\VF \in \tilde{\con}'_{(e, V)}$. Furthermore, we have that $\XF \tdash_{(a,X)} (\{(d, U)\} \cup \UF)$, $\UF \tilde{H}_{(d, U)} (\{(e, V)\} \cup\VF)$ and $\VF \tdash'_{(e, V)} \FF$, as requested.
 \end{pf}
 
 Set $\TTT(\HH) = \tilde{\HH}$.
 
 \begin{pn}\label{pn-funcT}
 $\fun{\TTT}{\CIF}{\SIF}$ is a functor.
 \end{pn}

As said, $\SIF$ is a full subcategory of $\CIF$. Let $\fun{\KKK}{\SIF}{\CIF}$ be the inclusion functor. In the remainder of this section we will show that the functors $\TTT$ and $\KKK$ establish an equivalence between the categories $\SIF$ and $\CIF$. 

Let to this end $\bA$ be a continuous information frame. For $i \in A$, $X \in \con_{i}$, $(k, K) \in \tilde{A}$,  $\XF \in \tilde{\con}_{(k, K)}$, $a \in  A$ and $(b, Y) \in \tilde{A}$
define
\begin{gather*}
X M^{\bA}_{i} (b, Y) \Leftrightarrow X \vDash_{i} (\{b\} \cup Y), \\
\XF N^{\bA}_{(k, K)} a \Leftrightarrow \XF \tdash_{(k, K)} (a, \{a\}).
\end{gather*}
Set $M_{\bA} = (M^{\bA}_{i})_{i \in A}$ and $N_{\bA} = (N^{\bA}_{(k, K)})_{(k, K)}\in \tilde{A}$.
 
\begin{lem}\label{lem-isoT}
\begin{enumerate}

\item\label{lem-isoT-1}
$\appmapf{M_{\bA}}{\bA}{\TTT(\bA)}$.

\item\label{lem-isoT-2}
$\appmapf{N_{\bA}}{\TTT(\bA)}{\bA}$.

\item\label{lem-isoT-3}
$M_{\bA} \circ N_{\bA} = \Id_{\bA}$.

\item\label{lem-isoT-4}
$N_{\bA} \circ M_{\bA} = \Id_{\TTT(\bA)}$.

\item\label{lem-isoT-5}
Let $\bA'$ be a further continuous information frame and $\appmapf{\HH}{\bA}{\bA'}$. Then
\[
M_{\bA} \circ \tilde{\HH} = \HH \circ M_{\bA'}.
\]
\end{enumerate}
\end{lem} 
\begin{pf}
\eqref{lem-isoT-1} and  \eqref{lem-isoT-2} follow with the corresponding properties of the entailment families.

\eqref{lem-isoT-3}
Let $a, i \in A$ and $X \in \con_{i}$. Then
\begin{align*}
X (M_{\bA} &\circ N_{\bA})_{i} a \\
\Leftrightarrow\mbox{} & (\exists (d, U) \in \tilde{A}) (\exists \UF \in \tilde{\con}_{(d, U)})\, X M^{\bA}_{i} (\{(d, U)\} \cup \UF) \kand \UF N^{\bA}_{(d, U)} a \\
\Leftrightarrow\mbox{}  &(\exists (d, U) \in \tilde{A}) (\exists \UF \in \tilde{\con}_{(d, U)}) (\forall (c, Z) \in \UF \cup \{(d, U)\})\, X \vDash_{i} (\{c\} \cup Z) \kand\mbox{} \\
&(\exists (e, V) \in \UF \cup \{(d, U)\})\, V \vDash_{e} a \\
\Leftrightarrow\mbox{}  & (\exists e \in A) (\exists V \in \con_{e})\, X \vDash_{i} (\{e\} \cup V) \kand V \vDash_{e} a \\
\Leftrightarrow \mbox{} &X \vDash_{i} a,
\end{align*}
where the left-to-right implication of the last equivalence follows with Lemma~\ref{lem-strong6} and the reverse implication is a consequence of Condition~\ref{dn-infofr}\eqref{dn-infofr-11}. The right-to-left implication of the next-to-last equivalence follows by setting $(d, U) = (e, V)$ and $\UF = \{(e, V)\}$.

\eqref{lem-isoT-4}
Let $(i, X), (b, Y) \in \tilde{A}$ and $\XF \in \tilde{\con}_{(i, X)}$. Assume that 
\[
\XF (N_{\bA} \circ M_{\bA})_{(i, X)} (b, Y).
\]
 Then there are $d \in A$ and $U \in \con_{d}$ so that $\XF N^{\bA}_{(i, X)}(\{d\} \cup U)$ and $U M^{\bA}_{d} (b, Y)$. 

From $\XF N^{\bA}_{(i, X)}(\{d\} \cup U)$ it follows that for every $a \in U \cup \{d\}$ there is some $(c_{a}, Z_{a}) \in \XF \cup \{(i, X)\}$ with $Z_{a} \vDash_{c_{a}} a$. Because $\XF \in \tilde{\con}_{(i, X)}$ we have for $(c_{a}, Z_{a}) \in \XF$ that $X \vDash_{i} (\{c_{a}\} \cup Z_{a})$. Hence, $X \vDash_{i} a$, which also holds if $(c_{a}, Z_{a}) = (i, X)$. So, we obtain that $X \vDash_{i} (\{d\} \cup U)$. 

As a consequence of  $U M^{\bA}_{d} (b, Y)$, we have that  $U \vDash_{d} (\{b\} \cup Y)$. From both we obtain that $X \vDash_{i} (\{b\} \cup Y)$. It follows that $\XF \tdash_{(i, X)} (b, Y)$.

Now, conversely, suppose that $\XF \tdash_{(i, X)} (b, Y)$. Then there is some $(c, Z) \in \XF \cup \{(i, X)\}$ with $Z \vDash_{c} (\{b\}\cup Y)$. By Interpolation there are thus $d \in A$ and $U \in \con_{d}$ with $Z \vDash_{c} (\{d\} \cup U)$ and $U \vDash_{d} (\{b\} \cup Y)$.  With the first property we obtain that $\XF N^{\bA}_{(i, X)} (\{d\} \cup U)$ and with the other one that $U M^{\bA}_{d} (b, Y)$. So, we have that $\XF (N_{\bA} \circ M_{\bA})_{(i, X)} (b, Y)$.

\eqref{lem-isoT-5}
Let $i \in A$, $X \in \con_{i}$ and $(b, Y) \in \tilde{A'}$. Then we have with Condition~\ref{dn-am}\eqref{dn-am-4} and Lemma~\ref{pn-amint-1} that
\begin{align}
X (&M_{\bA} \circ \tilde{\HH})_{i} (b, Y) \notag \\
\Leftrightarrow\mbox{} &(\exists  (d, U) \in \tilde{A}) (\exists \UF \in \tilde{\con}_{(d, U)})\, X M^{\bA}_{i} (\{(d, U)\} \cup \UF) \kand \UF \tilde{H}_{(d, U)} (b, Y) \notag\\
\Leftrightarrow\mbox{} &(\exists  (d, U) \in \tilde{A}) (\exists \UF \in \tilde{\con}_{(d, U)}) (\forall (c, Z) \in \UF \cup \{(d, U)\})\, X \vDash_{i} (\{c\} \cup Z) \kand\mbox{} \notag\\
&(\exists (e, V) \in \UF \cup \{(d, U)\})\, V H_{e} (\{b\} \cup Y) \notag\\
\Leftrightarrow\mbox{} &X H_{i} (\{b\} \cup Y) \label{eq-isoT-5-1} \\
\Leftrightarrow\mbox{} & (\exists k \in A') (\exists K \in \con'_{k})\, X H_{i} (\{k\} \cup K) \kand K \vDash'_{k} (\{b\} \cup Y) \label{eq-isoT-5-2}\\
\Leftrightarrow\mbox{} &(\exists k \in A') (\exists K \in \con'_{k})\,  X H_{i} (\{k\} \cup K) \kand K M^{\bA'}_{k} (b, Y)  \notag \\
\Leftrightarrow\mbox{} &X (\HH \circ M_{\bA'})_{i} (b, Y), \notag
\end{align}
where in \eqref{eq-isoT-5-1} the left-to-right implication follows with Lemma~\ref{lem-amstrong3}. In the reverse direction the existence of $e \in A$ and $V \in \con_{e}$ is a consequence of Lemma~\ref{pn-amint} and the choice of $(d, U) = (e, V)$ and $\UF = \{(e, V)\}$. Similarly, in \eqref{eq-isoT-5-2}, for the left-to-right implication Lemma~\ref{pn-amint} is applied and the reverse one follows with Conditions~\ref{dn-am}\eqref{dn-am-1} and \eqref{dn-am-4}.
\end{pf}
 
 For $\bA \in \CIF$ set $\kappa_{\bA} = M_{\bA}$ and for $\bA \in \SIF$ define $\sigma_{\bA} = M_{\bA}$.
 
 \begin{pn}\label{pn-iso-ks}
 $\fun{\kappa}{\ID_{\CIF}}{\KKK \circ \TTT}$ and $\fun{\sigma}{\ID_{\SIF}}{\TTT \circ \KKK}$ are natural isomorphisms.
 \end{pn}
 
 \begin{thm}\label{thm-eqrefl}
 \begin{enumerate}
 
 \item\label{thm-eqrefl-1}
 The category $\SIF$ is a full reflective subcategory of $\CIF$.
 
 \item\label{thm-eqrefl-2}
 The categories $\SIF$ and  $\CIF$ are equivalent.
 \end{enumerate}
 \end{thm}
 
Finally, let us summarise what we have achieved in this and the previous section.
 
 \begin{cor}\label{cor-eqdcat}
 The categories $\SIF$, $\CIF$, $\CIS$ and $\SCIS$ are all equivalent.
 \end{cor}

In the next section we will investigate the extension of continuous information frames by a truth element. We will now study whether the property of a continuous information frame to have a truth element and the property of an approximating family to respect truth elements are preserved under the functor $\TTT$.

\begin{pn}\label{pn-tpres1}
Let $\bA$ be a continuous information frame with truth element $\bt$. Then $(\bt, \emptyset)$ is a truth element of $\TTT(\bA)$. 
\end{pn}
\begin{pf}
We have to show that $(\bt, \emptyset)$ satisfies Condition~\eqref{cn-T}. Let $(a, X) \in \tilde{A}$. Since $\bt$ is a truth element of $\bA$, it follows with Condition~\eqref{cn-T} and Weakening that $X \vDash_{a} \bt$. Hence, $\emptyset \tdash_{(a, X)} (\bt, \emptyset)$.
\end{pf}

\begin{pn}\label{pn-tpres2}
Let $\bA$, $\bA'$ be continuous information frames with truth elements $\bt$ and $\bt'$, respectively, and $\appmapf{\HH}{\bA}{\bA'}$ so that truth elements are respected. Then $\TTT(\HH)$ respects the truth elements $(\bt, \emptyset)$ and $(\bt', \emptyset)$.
\end{pn}
\begin{pf}
Since $\HH$ respects the truth elements $\bt$ and $\bt'$, we have that $\emptyset H_{\bt} \bt'$. It follows that $\emptyset \tilde{H}_{(\bt, \emptyset)} (\bt', \emptyset)$, as required.
\end{pf}

Let $\CIF_{\bt}$ and $\SIF_{\bt}$, respectively,  be the categories of continuous and strong continuous information frames with truth elements and truth element respecting approximating families. 

 \begin{thm}\label{thm-eqreflt}
 \begin{enumerate}
 
 \item\label{thm-eqreflt-1}
 The category $\SIF_{\bt}$ is a full reflective subcategory of $\CIF_{\bt}$.
 
 \item\label{thm-eqreflt-2}
 The categories  $\SIF_{\bt}$ and $\CIF_{\bt}$ are equivalent.
 \end{enumerate}
 \end{thm}

\section{Truth elements}\label{sec-truth}

We will show in this section that any continuous information frame can be extended by a truth element. 

Let $\bA = (A, (\con_{i})_{i \in A}, (\vDash_{i})_{i \in A})$ be a continuous information frame and $\bt \notin A$. Set
\begin{align*}
&\ov{A} = A \cup \{\bt\},					\\
&\ov{\con}_{a} = \begin{cases}
				\con_{a} \cup \set{X \cup \{\bt\}}{X \in \con_{a}}  & \text{if $a \in A$,}					\\
				\{\{\bt\}, \emptyset\}						& \text{if $a = \bt$,}
			\end{cases}				\tag{$a \in \ov{A}$}\\
&X \odash_{a} c \Leftrightarrow
				\begin{cases}
				X \setminus \{\bt\} \vDash_{a} c				& \text{if $a \neq \bt$ and $c \neq \bt$,}	\\
				c = \bt								& \text{otherwise.}					\tag{$a \in \ov{A}, X \in \ov{\con}_{a}$}
				\end{cases}
\end{align*}
				
and define
\[
\WWW(\bA) = (\ov{A}, (\ov{\con}_{i})_{i \in \ov{A}}, (\odash_{i})_{i \in \ov{A}}).
\]

\begin{thm}\label{thm-exstr}
$\WWW(\bA)$ is a continuous information frame with truth element $\bt$.
\end{thm}
\begin{pf}
We have to verify Condition~\eqref{cn-T} and the conditions in Definition~\ref{dn-infofr}. We only consider Condition~\ref{dn-infofr}\eqref{dn-infofr-11}. The other conditions follow easily from the definition.

Let to this end $i \in \ov{A}$, $X \in \ov{\con}_{i}$ and $Y \fsubset \ov{A}$ with $X \odash_{i} Y$. If $Y = \{\bt\}$, set $c = \bt$ and $Z = \emptyset$. Then $X \odash_{i} (\{c\} \cup Z)$ and 
$Z \odash_{c} Y$. In the other case, $i \neq \bt$, Let $X' = X \setminus \{\bt\}$ and $Y' = Y \setminus \{\bt\}$. Then $X' \vDash_{i} Y$. Hence, there exist $c \in A$ and $Z \in \con_{c}$ with $X' \vDash_{i} (\{c\} \cup Z)$ and $Z \vDash_{c} Y'$. It follows that $X \odash_{i} (\{c\} \cup Z)$ and $Z \odash_{c} Y'$. As always $Z \odash_{c} \bt$, we also have that $Z \odash_{c} Y$, if $\bt \in Y$.  Otherwise, $Y = Y'$.  
\end{pf}

Let $\bA'$ be another continuous information frame and $\appmapf{\HH}{\bA}{\bA'}$. We will now extend $\HH$ to an approximating family from $\WWW(\bA)$ to $\WWW(\bA')$ so that  truth elements are respected. Let to this end $\bt' \notin A$ such that $\ov{A'} = A' \cup \{\bt'\}$. For $a \in \ov{A}$, $X \in \ov{\con}_{a}$ and $c \in \ov{A}'$ set
\[
X \ov{H}_{a} c \Leftrightarrow 
				\begin{cases}
				X \setminus \{\bt\} H_{a} c & \text{if $a \neq \bt$ and $c \neq \bt'$,} \\
				c = \bt'				& \text{otherwise.}
				\end{cases}				
\]

\begin{pn}\label{pn-extm}
\begin{enumerate}

\item\label{pn-extm-1}
$\appmapf{\ov{\HH} = (\ov{H}_{a})_{a \in \ov{A}}}{\WWW(\bA)}{\WWW(\bA')}$.

\item\label{pn-extm-2}
$\ov{\HH}$ respects the truth elements $\bt$ and $\bt'$.

\end{enumerate}
\end{pn}
\begin{pf}
The proof is as in the case of Theorem~\ref{thm-exstr}.
\end{pf}

Set $\WWW(\HH)  = \ov{\HH}$.

\begin{pn}\label{pn-functW}
$\fun{\WWW}{\CIF}{\CIF_{\bt}}$ is a functor.
\end{pn}

$\CIF_{\bt}$ is a subcategory of $\CIF$. Let $\fun{\VVV}{\CIF_{\bt}}{\CIF}$ be the inclusion functor. As we will show next, the functors $\WWW$ and $\VVV$ establish an equivalence between $\CIF_{\bt}$ and $\CIF$. Let to this end $\bA$ be continuous information frame. For $i, a \in A$, $X \in \con_{i}$, $b, c \in \ov{A}$ and $Z \in \ov{\con}_{b}$ define
\begin{gather*}
X J^{\bA}_{i} c \Leftrightarrow X \odash_{i} c, \\
Z L^{\bA}_{b} a \Leftrightarrow Z \odash_{b} a.
\end{gather*}

Set $J_{\bA} = (J^{\bA}_{i})_{i \in A}$ and $L_{\bA} = (L^{\bA}_{b})_{b \in \ov{A}}$.

\begin{lem}\label{lem-JL}
\begin{enumerate}

\item\label{lem-JL-1}
$\appmapf{J_{\bA}}{\bA}{\WWW(\bA)}$.

\item\label{lem-JL-2}
$\appmapf{L_{\bA}}{\WWW(\bA)}{\bA}$.

\item\label{lem-JL-3}
$J_{\bA} \circ L_{\bA} = \Id_{\bA}$.

\item\label{lem-JL-4}
$L_{\bA} \circ J_{\bA} = \Id_{\WWW(\bA)}$.

\item\label{lem-JL-5}
Let $\bA'$ be a further continuous information frame and $\appmapf{\HH}{\bA}{\bA'}$, Then
\[
J_{\bA} \circ \ov{\HH} = \HH \circ J_{\bA'}.
\]

\end{enumerate}
\end{lem}
\begin{pf}
\eqref{lem-JL-1} and \eqref{lem-JL-2} follow again with corresponding properties of $(\odash_{i})_{i \in \ov{A}}$. Moreover, \eqref{lem-JL-3} and \eqref{lem-JL-4} are consequences of Condition~\ref{dn-infofr}\eqref{dn-infofr-11} and Lemma~\ref{lem-strong6}. We only consider Statement~\eqref{lem-JL-5}.

Let $i \in A$, $X \in \con_{i}$ and $c \in \ov{A'}$.  Then it follows with Lemmas~\ref{pn-amint}, \ref{lem-amstrong3} and Condition~\ref{dn-am}\eqref{dn-am-1} that
\begin{align*}
X (\HH \circ J_{\bA'})_{i} c 
&\Leftrightarrow (\exists e \in A) (\exists E \in  \con_{e})\, X H_{i} (\{e\} \cup E) \kand E J^{\bA'}_{e} c \\
&\Leftrightarrow (\exists e \in A) (\exists E \in  \con_{e})\, X \ov{H}_{i} (\{e\} \cup E) \kand E \odash'_{e} c \\
&\Leftrightarrow X \ov{H}_{i} c \\
&\Leftrightarrow (\exists k \in \ov{A}) (\exists K \in \ov{\con}_{k})\, X \odash_{i} (\{k\} \cup K) \kand K \ov{H}_{k} c \\
&\Leftrightarrow (\exists k \in \ov{A}) (\exists K \in \ov{\con}_{k})\, X J^{\bA}_{i} (\{k\}\cup K) \kand K \ov{H}_{k} c \\
&\Leftrightarrow X (J_{\bA} \circ \ov{\HH})_{i} c.
\end{align*}
\end{pf}

For $\bA \in \CIF$ set $\theta_{\bA} = J_{\bA}$ and for $\bA \in \CIF_{\bt}$ define $\rho_{\bA} = J_{\bA}$.

\begin{pn}\label{pn-isotr}
$\fun{\theta}{\ID_{\CIF}}{\VVV \circ \WWW}$ and $\fun{\rho}{\ID_{\CIF_{\bt}}}{\WWW \circ \VVV}$ are natural ismorphisms.
\end{pn}

\begin{thm}\label{thm-eqbt}
\begin{enumerate}

\item\label{thm-eqbt-1}
The category $\CIF_{\bt}$ is a reflective subcategory of $\CIF$.

\item\label{thm-eqbt-2}
The categories $\CIF$ and $\CIF_{\bt}$ are equivalent.

\end{enumerate}
\end{thm}

\begin{cor}\label{cor-t}
\begin{enumerate}

\item\label{cor-t-1}
The category $\SIF_{\bt}$ is a reflective subcategory of $\CIF$.

\item\label{cor-t-2}
The categories $\SIF_{\bt}$ and $\CIF$ are equivalent.

\end{enumerate}
\end{cor}

\section{Stratified conjunctive logics}\label{sec-log}

As we have seen so far, a continuous information frame is, in particular, a graph where each node is associated with a rudimentary logic consisting of a consistency predicate that specifies which finite subsets of atomic statements are consistent, and an entailment relation that specifies which atomic statements follow from such consistent subsets. In this section we will study the logic generated by such frames.

\begin{dn}\label{dn-form}
Let $P$ be a set, the elements of which are called \emph{atomic propositions}, and $\top \in P$ be a syntactic constant for ``true''. The set  $\LLL(P)$ of \emph{formulae} is defined in the usual inductive way:
\begin{enumerate}

\item Every atomic proposition is a formula.

\item If $\varphi$ and $\psi$ are formulae, so is $\varphi \land \psi$.

\end{enumerate}
\end{dn}

Given a non-empty finite subset $\Gamma = \{\varphi_{1}, \varphi_{2}, \ldots, \varphi_{n} \}$ of $\LLL(P)$,  the formula $\varphi_{1} \land \varphi_{2} \land \cdots \land \varphi_{n}$ with bracketing to the left is abbreviated as $\bigwedge \Gamma$. In case $\Gamma = \emptyset$, let $\bigwedge \Gamma = \top$.

For any $\varphi \in \LLL(P)$, 
\[
\overline{\varphi} = \{p_{1}, p_{2}, \ldots, p_{n} \},
\]
where $p_{1}, p_{2}, \ldots, p_{n}$ are all the atomic propositions that occur in $\varphi$. Similarly,  
\[
\overline{\Delta}= \bigcup\set{\overline{\varphi}}{\varphi \in \Delta}
\]
for any non-empty subset $\Delta$ of $\LLL(P)$. In case $\Delta = \emptyset$, set $\ov{\Delta} = \top$. 
Then $\bigwedge \colon \powf{P} \leftrightarrows \LLL(P) : \ov{\,\cdot\,}$ is an embedding-retraction.
If there is no ambiguity, the singleton $\{\varphi\}$ is abbreviated as $\varphi$.

\begin{dn}\label{dn-stratlog}
A \emph{stratified conjunctive logic} $\PP = (P, (P_{p})_{p \in P}, (\vdash^{p})_{p \in P})$ consists of a set $P$ of atomic propositions so that there is a syntactic constant $\top \in P$ for ``true'', a family $(P_{p})_{p \in P}$ of subsets $P_{p}$ of $P$, and a family $(\vdash^{p})_{p \in P}$ of relations $\vdash^{p} \subseteq (\powf{\LLL(P_{p})} \cup \{\{p\}\}) \times \LLL(P_{p})$ such that for $p, q \in P$, $\Gamma \in \powf{\LLL(P_{p})} \cup \{\{p\}\}$, $\Delta \in \powf{\LLL(P_{p})}$ and $\varphi, \psi, \theta \in \LLL(P_{p})$,
\begin{enumerate}

\item\label{dn-stratlog-0}
$q \in P_{p} \Rightarrow p \vdash^{p} q$.
 
 \item\label{dn-stratlog-1}
 $q \vdash^{q} p \Rightarrow P_{p} \subseteq P_{q}$,
 
 \item \label{dn-stratlog-2}
 $q \vdash^{q} p \kand \Gamma \vdash^{p} \theta \Rightarrow \Gamma \vdash^{q} \theta$,
 
 \item \label{dn-stratlog-3}
 The relation $\vdash^{p}$ is closed under the Rules (R$\top$), (L$\land$), (R$\land$), (Cut) and (W):
 \[
\begin{array}{cc}
\multicolumn{2}{c}{
\dfrac{}{\Gamma \vdash^{p} \top}\,\, \text{(R$\top$)} } \\[3ex]
\ddoublefrac{\Delta, \varphi, \psi \vdash^{p} \theta}{\Delta, \varphi \land \psi \vdash^{p} \theta}\,\, \text{(L$\land$)} &
\ddoublefrac{\Gamma \vdash^{p} \varphi\,\,\,\, \Gamma \vdash^{p} \psi}{\Gamma \vdash^{p} \varphi \land \psi}\,\,  \text{(R$\land$)} \\[3ex]
 \dfrac{\Gamma \vdash^{p} \varphi\,\,\,\, \varphi \vdash^{p} \psi}{\Gamma \vdash^{p} \psi}\,\, \text{(Cut)} &
 \dfrac{\Delta \vdash^{p}  \varphi,\,\,\, \xi \in \LLL(P_{p})}{\Delta, \xi \vdash^{p} \varphi}\,\, \text{(W)}
\end{array}
\]
where the double line indicates that the rule can be used in both directions.
\end{enumerate}
\end{dn}

The `backwards' rules in (L$\land$) and (R$\land$) are not present in usual sequent calculus where they are consequences of the cut and the identity rule.  The latter, however, is not used here. Note that on the side of structural rules only weakening, that is (W), is mentioned. Exchange and contraction are kept implicit. Thus we are working with sets rather than sequences.

\begin{dn}\label{dn-stratlogcont}
A stratified conjunctive logic $\PP$ is called \emph{continuous}, if, in addition, Rule (INT) holds, for all $p \in P$, $\Gamma \in \powf{\LLL(P_{p})} \cup \{\{p\}\}$ and $\varphi \in \LLL(P_{p})$:
\begin{equation}
\Gamma \vdash^{p} \varphi \Rightarrow (\exists r \in P_{p}) (\exists \xi \in  \LLL(P_{r}))\, \Gamma \vdash^{p} r \land \xi\, \kand\, \xi \vdash^{r} \varphi. \tag{INT}
\end{equation}
\end{dn}

\begin{lem}\label{lem-rulebarwedge}
Let $p \in P$,  $\Gamma \in \powf{\LLL(P_{p})} \cup \{p\}$ and $\varphi \in \LLL(P)$. Then the following rules are derivable:
\[
\begin{array}{cc}
\ddoublefrac{\Gamma \vdash^{p} \varphi}{\ov{\Gamma}  \vdash^{p} \varphi} \qquad
& \ddoublefrac{\Gamma \vdash^{p} \varphi}{\Gamma \vdash^{p} \bigwedge \ov{\varphi}}
\end{array}
\]
\end{lem}
\begin{pf}
We only consider the left rule. The other rule follows similarly.

For $\nu = 0, \ldots, \|\Gamma\| -1$, we derive rules
\begin{equation}\label{eq-barrule}
\ddoublefrac{\ov{\Gamma_{\nu}}, \Gamma'_{\nu}, \psi \vdash^{p} \varphi}{\ov{\Gamma_{\nu}} \cup \ov{\psi}, \Gamma'_{\nu}  \vdash^{p} \varphi},
\end{equation}
where $\Gamma_{0} = \emptyset$ and $\Gamma'_{0}, \psi = \Gamma$. Rule~\eqref{eq-barrule} follows by induction on the built-up of $\psi$ using (L$\land$). If $\Gamma'_{\nu} = \emptyset$, we are done. Otherwise, $\Gamma'_{\nu} = \Gamma'_{\nu+1}, \theta$. Then we set $\Gamma_{\nu+1} = \Gamma_{\nu} \cup \ov{\psi}$ and repeat the procedure.
\end{pf}

By applying the left rule twice, from top to bottom and vice versa, we get the following useful rule:
\begin{equation}\label{rule-bar}
\ddoublefrac{\Gamma \vdash^{p} \varphi}{\bigwedge \ov{\Gamma}  \vdash^{p} \varphi}
\end{equation}

Note that as a consequence of the interpolation rule (INT), the converse of the cut rule also holds for each of the relations $\vdash^{p}$.

\begin{lem}\label{lem-invcut}
Let $\PP$ be a continuous stratified conjunctive logic. Then for $\Gamma \in \powf{\LLL(P_{p})} \cup \{\{p\}\}$ and $\psi \in  \LLL(P_{p})$ there is some $\varphi \in \LLL(P_{p})$ so that
\[
\dfrac{\Gamma \vdash^{p} \psi}{\Gamma \vdash^{p} \varphi\,\,\, \varphi \vdash^{p} \psi}.
\]
\end{lem}
\begin{pf}
Assume that $\Gamma \vdash^{p} \psi$. Then, by Rule (INT),  there exist $q \in P_{p}$ and $\varphi \in \LLL(P_{q})$ so that $\Gamma \vdash^{p} q \land \varphi$ and $\varphi \vdash^{q} \psi$. It follows that $\Gamma \vdash^{p} q$ and $\Gamma \vdash^{p} \varphi$.
As a consequence of Lemma~\ref{lem-rulebarwedge} we obtain with (L$\land$) in the first case that $\bigwedge \ov{\Gamma} \vdash^{p} q$.

By our assumption on $\Gamma$ we know that $\Gamma = \{p\}$ or $\Gamma \fsubset \LLL(P_{p})$. If $\Gamma = \{p\}$, we have that $p \vdash^{p} q$. In the other case it follows with  \ref{dn-stratlog}\eqref{dn-stratlog-0} that for all $r \in \ov{\Gamma}$, $p \vdash^{p} r$ and hence with  (R$\land$)  that $p \vdash^{p} \bigwedge \ov{\Gamma}$. With Cut it follows that in this case too $p \vdash^{p} q$.
  Thus,  $\varphi \vdash^{p} \psi$, because of \ref{dn-stratlog}\eqref{dn-stratlog-2}.
\end{pf}

\begin{lem}\label{lem-sint}
Let $\PP$ be a stratified conjunctive logic. Then Rule (INT) is equivalent to its simplification (SINT), which says that  for all $p \in P$, $\Gamma \in \powf{\LLL(P_{p})} \cup \{\{p\}\}$ and $\varphi \in \LLL(P_{p})$:
\begin{equation}
\Gamma \vdash^{p} \varphi \Rightarrow (\exists r \in P_{p})\, \Gamma \vdash^{p} r\, \kand\, r \vdash^{r} \varphi. \tag{SINT}
\end{equation}
\end{lem}
\begin{pf}
Assume that (INT) holds and $\Gamma \vdash^{p} \varphi$. Then there are $r \in P_{p}$ and $\xi \in \LLL(P_{r})$ with $\Gamma \vdash^{p} r$, $\Gamma \vdash^{p} \xi$ and $\xi \vdash^{r} \varphi$. Since $\xi \in \LLL(P_{r})$, it follows as in the proof of Lemma~\ref{lem-invcut} that $r \vdash^{r} \varphi$.
 
For the reverse direction assume that $\Gamma \vdash^{p} \varphi$, Then there is some $r \in P_{p}$ with $\Gamma \vdash^{p} r$ and $r \vdash^{r} \varphi$. Now apply Rule (SINT) to 
$r \vdash^{r} \varphi$.  Then there is some $q \in P_{r}$ with $r \vdash^{r} q$ and $q \vdash^{q} \varphi$. As in the proof of Lemma~\ref{lem-invcut}, $\Gamma \vdash^{p} r$ implies that $p \vdash^{p} r$.  With \ref{dn-stratlog}\eqref{dn-stratlog-2}  we thus obtain that $r \vdash^{p} q$ and $q \vdash^{r} \varphi$. Hence, we have that $\Gamma \vdash^{p} r \land q$ and $q \vdash^{r} \varphi$. Set $\xi = q$.  Then $\xi \in \LLL(P_{r})$. 
\end{pf}

\begin{dn}\label{dn-glob}
Let $\PP'$ be a further stratified conjunctive logic. A \emph{global consequence relation} $\corel$ from $\PP$ to $\PP'$, written $\cormap{\corel}{\PP}{\PP'}$, is a family $(\corel^{p})_{p \in P}$ of relations $\corel^{p} \subseteq (\powf{\LLL(P_{p})} \cup \{\{p\}\}) \times \LLL(P')$ such that
\begin{enumerate}

\item\label{dn-glob-1}
For every $p \in P$, $\corel^{p}$ is closed under the rules (R$\top$), (W), (L$\land$), (R$\land$) as well as (Lcut) and (Rcut):
\[
\begin{array}{cc}
\dfrac{\Gamma \vdash^{p} \varphi \,\,\,\, \varphi \corel^{p} \psi}{\Gamma \corel^{p} \psi}\,\, \text{(Lcut)} &
\dfrac{\Gamma \corel^{p} q \land \psi\,\,\,\, \psi \vdash'^{q} \theta}{\Gamma \corel^{p} \theta}\,\, \text{(Rcut)}, 
\end{array}
\]
where $\Gamma \in \powf{\LLL(P_{p})} \cup \{\{p\}\}$, $\varphi \in \LLL(P_{p})$, $\psi \in \LLL(P'_{q}) \cup \{q\}$ and $\theta \in \LLL(P'_{q})$

\item\label{dn-glob-2}
For all $p, r \in P$ with $p \vdash^{p} r$, if $\Gamma \corel^{r} \psi$ then $\Gamma \corel^{p} \psi$.

\item\label{dn-glob-3}
For all $p \in P$, $\Gamma \in \powf{\LLL(P_{p})} \cup \{\{p\}\}$ and $\varphi \in \LLL(P')$,
\begin{multline*}
 \Gamma \corel^{p} \varphi \Rightarrow (\exists r \in P) (\exists \theta \in \LLL(P_{r}) \cup \{r\})\\
  (\exists q \in P') (\exists \psi \in \LLL(P'_{q}) \cup \{q\})\, \Gamma \vdash^{p} r \land \theta\,\,\,\,  \theta \corel^{r} q \land \psi\, \kand\, \psi \vdash'^{q} \varphi.
 \end{multline*}

\end{enumerate}
\end{dn}

Note in the last requirement that since $\Gamma \in \powf{\LLL(P_{p})} \cup \{\{p\}\}$, we have that $p \vdash^{p} \bigwedge \ov{\Gamma}$. Because of $\Gamma \vdash^{p} r \land \theta$, it follows that $\Gamma \vdash^{p} r$. Hence, $p \vdash^{p} r$, that is, $r \in P_{p}$. 

As for the family members $\corel^{p}$ of a global consequence relation the same rules hold as for the relations of a stratified conjunctive logic, similar rules as in Lemma~\ref{lem-rulebarwedge} and in \eqref{rule-bar} can be derived.

\begin{lem}\label{lem-globprop}
Let $\PP$ and $\PP'$ be continuous stratified conjunctive logics and $\protect{\corel} \colon \PP \bowtie \PP'$. Then the following rules hold for all $p, r \in P$, $\Gamma \in \powf{\LLL(P_{p})} \cup \{\{p\}\}$, $\theta \in \LLL(P_{p})$ and $\varphi \in \LLL(P')$:
\[
\Gamma \vdash^{p} r \land \theta \kand \theta \corel^{r} \varphi \Rightarrow \Gamma \corel^{p} \varphi.
\]
\end{lem}
\begin{pf}
The statement follows with \ref{dn-glob}\eqref{dn-glob-2} and (Lcut).
\end{pf}

\begin{lem}\label{lem-equivglob3}
Let $\PP$ and $\PP'$ be continuous stratified conjunctive logics. Then for any family $(\corel^{p})_{p \in P}$ with $\corel^{p} \subseteq (\powf{\LLL(P_{p})} \cup \{\{p\}\}) \times \LLL(P')$, 
Condition~\ref{dn-glob}\eqref{dn-glob-3} is equivalent to the following Conditions~\eqref{lem-equivglob3-1} and \eqref{lem-equivglob3-2}:
\begin{enumerate}

\item\label{lem-equivglob3-1}
There are $s \in P$ and $\xi \in \LLL(P_{s}) \cup \{s\}$ so that, if $\Gamma \corel^{p} \varphi$ then $\Gamma \vdash^{p} s \land \xi$ and $\xi \corel^{s} \varphi$.

\item \label{lem-equivglob3-2}
There are $q \in  P'$ and $\psi \in \LLL(P'_{q}) \cup \{q\}$ so that, if $\Gamma \corel^{p} \varphi$ then $\Gamma \corel^{p} q \land \psi$ and $\psi \vdash'^{q} \varphi$.

\end{enumerate}
\end{lem}
\begin{pf}
\eqref{lem-equivglob3-1}  and \eqref{lem-equivglob3-2} are a consequence of \ref{dn-glob}\eqref{dn-glob-3} and (Rcut) or (Lcut).  Conversely, to derive \ref{dn-glob}\eqref{dn-glob-3}, assume  $\Gamma \corel^{p} \varphi$ and apply \eqref{lem-equivglob3-1} to obtain $s \in P$ and $\xi \in \LLL(P_{s}) \cup \{s\}$ with  $\Gamma \vdash^{p} s \land \xi$ and $\xi \corel^{s} \varphi$ and then apply \eqref{lem-equivglob3-2} to  $\xi \corel^{s} \varphi$.
\end{pf}

For every  continuous stratified conjunctive logic  $\PP$, the family $(\vdash^{p})_{p \in P}$ is a global consequence relation, the identity $\Id_{\PP}$ on $\PP$.

Let  $\PP$, $\PP'$ and $\PP''$ be  continuous stratified conjunctive logics, $\cormap{\corel_{1}}{\PP}{\PP'}$ and $\cormap{\corel_{2}}{\PP'}{\PP''}$. For $p \in P$ set
\begin{multline*}
(\corel_{1} \circ \corel_{2})^{p} = \{\, (\Gamma, \varphi) \mid \Gamma \in \powf{\LLL(P_{p})} \cup \{\{p\}\} \kand \varphi \in \LLL(P'') \kand \mbox{} \\
 (\exists q \in P') (\exists \psi \in \LLL(P'_{q}) \cup \{q\}) \Gamma \corel^{p}_{1} q \land \psi\, \kand\, \psi \corel_{2}^{q} \varphi \,\}
\end{multline*}
and define
\[
\corel_{1} \circ \corel_{2} = ((\corel_{1} \circ \corel_{2})^{p})_{p \in P}.
\]
  
\begin{lem}\label{lem-corcirc}
Let  $\PP$, $\PP'$ and $\PP''$ be  continuous stratified conjunctive logics, $\corel \colon \PP \bowtie \PP'$ and $\cormap{\corel'}{\PP'}{\PP''}$.
Then the following statements hold:
\begin{enumerate}

\item\label{lem-corcirc-1}
$(\corel \circ \corel'): \PP \bowtie \PP''$.

\item\label{lem-corcirc-2}
$\mathrel{(\vdash^{p})_{p \in P}} \circ \corel\, =\, \corel\, =\,  \corel \circ \mathrel{(\vdash'^{q})_{q \in P'}}$.

\end{enumerate}
 \end{lem}

Let $\CSL$ be the category of continuous stratified conjunctive logics and global consequence relations.

\section{Information frames and their stratified logics}\label{sec-inflog}

The goal of this section is to show that strong continuous information frames with a truth element and continuous stratified conjunctive logics are mutually dependent. We begin by showing how each strong continuous information frame is associated with a continuous stratified conjunctive logic.

Let to this end $\bA$ be a strong continuous information frame with truth element.  Set $P^{\bA} = A$ with $\top = \bt$ and let the set $\LLL(P^{\bA})$ of formulae be inductively generated as in Definition~\ref{dn-form} by starting with the elements of $A$ as atomic propositions. Then, if $X = \{ p_{1}, \ldots, p_{n} \}$ is a finite subset of $A$, $\bigwedge X = p_{1} \land \cdots \land p_{n}$, with bracketing to the left.

For $i \in P^{\bA}$, $\Gamma \fsubset \LLL(P^{\bA})$ and $\varphi \in \LLL(P^{\bA})$ , let
\begin{gather*}
P^{\bA}_{i} = \set{j \in A}{\{i\} \vDash_{i} j},  \\ 
\Gamma \vdash_{\bA}^{i} \varphi \Leftrightarrow \ov{\Gamma} \in \con_{i} \kand \ov{\Gamma} \vDash_{i} \ov{\varphi},
\end{gather*} 
where the sets $\ov{\varphi}$ and $\ov{\Gamma}$ are defined as in the previous section. 

Note that since $\bA$ is strong, we have
\begin{equation}\label{eq-logfr}
\Gamma \in \powf{\LLL(P^{\bA}_{i})} \cup \{\{i\}\} \Leftrightarrow \ov{\Gamma} \in \con_{i}.
\end{equation}
Set
\[
\PP^{\bA} = (P^{\bA}, (P^{\bA}_{i})_{i \in P^{\bA}}, (\vdash_{\bA}^{i})_{i \in P^{\bA}}).
\]

\begin{thm}\label{thm-frlog}
Let $\bA$ be a strong continuous information frame. Then $\PP^{\bA}$ is a continuous stratified conjunctive logic  such that for all $i \in A$:
\begin{enumerate}

\item\label{thm-frlog-1}
For all $\Gamma \in \powf{\LLL(P^{\bA}_{i})} \cup \{\{i\}\}$ and $\varphi \in \LLL(P^{\bA}_{i})$,
\[
\Gamma \vdash_{\bA}^{i} \varphi \Rightarrow \ov{\Gamma} \in \con_{i} \kand \ov{\Gamma} \vDash_{i} \ov{\varphi}.
\]

\item\label{thm-frlog-2}
For all $X \in \con_{i}$ and $Y \fsubset A$,
\[
X \vDash_{i} Y \Rightarrow  X \in \powf{\LLL(P^{\bA}_{i})} \cup \{\{i\}\} \kand X \vdash_{\bA}^{i} \bigwedge Y. 
\]
\end{enumerate}
\end{thm}
\begin{pf} 
The statements~\eqref{thm-frlog-1} and \eqref{thm-frlog-2} of the theorem are a consequence of \eqref{eq-logfr}.
To show that $\PP^{\bA}$ is a stratified conjunctive logic, the requirements in Definition~\ref{dn-stratlog} have to be verified. Conditions~\eqref{dn-stratlog-0}-\eqref{dn-stratlog-2} are obvious. For Condition~\eqref{dn-stratlog-3} note that Rules (RT) and (Cut) follow with the corresponding properties of the entailment relation.  Rules (L$\land$) and (R$\land$) obviously hold as $\ov{\varphi \land \psi} = \ov{\varphi} \cup \ov{\psi}$. For (W) let $\Delta \fsubset \LLL(P^{\bA}_{p})$ and $\psi \in \LLL(P^{\bA}_{p})$. Then $\{p\} \vDash_{p} \ov{\Delta}$ and $\{p\} \vDash_{p} \ov{\psi}$. Hence, $\{p\} \vDash_{p} (\ov{\Delta} \cup \ov{\psi})$, which implies that $\ov{\Delta} \cup \ov{\psi} \in \con_{p}$. Now, (W) follows with Property~\ref{dn-infofr}\eqref{dn-infofr-5}.

It remains to show that $\PP^{\bA}$ is continuous, that is, we have to prove that Rule (INT) holds. Assume that $\Gamma \vdash_{\bA}^{p} \varphi$. Then $\ov{\Gamma} \in \con_{p}$ and $\ov{\Gamma} \vDash^{p} \ov{\varphi}$. With Interpolation we obtain that there are $r \in P^{\bA}$ and $Z \in \con_{r}$ with $\ov{\Gamma} \vDash_{p} (\{r\} \cup Z)$ and $Z \vDash_{r} \ov{\varphi}$. It follows that $\{r\} \in \con_{p}$. Set $\xi = \bigwedge Z$. Since $\bA$ is strong, we obtain that $r \in P^{\bA}_{p}$ and $\xi \in \LLL(P^{\bA}_{r})$. Moreover, we have that $\Gamma \vdash_{\bA}^{p} r \land \xi$ and $\xi \vdash_{\bA}^{r} \varphi$.
\end{pf}

We want to construct a functor $\fun{\CCC}{\SIF_{\bt}}{\CSL}$. For $\bA \in \SIF_{\bt}$, set  $\CCC(\bA) = \PP^{\bA}$.
It remains to define how $\CCC$ acts on approximating families. Let to this end $\bA'$ be a further strong continuous information frame and $\appmapf{\HH}{\bA}{\bA'}$ truth element respecting. For  $\Gamma \in \powf{\LLL(A)$},  $\varphi \in \LLL(A')$ and $i \in A$ set 
\[
\Gamma \corel^{i}_{\HH} \varphi \Leftrightarrow \ov{\Gamma} \in \con_{i} \kand \ov{\Gamma} H_{i} \ov{\varphi}.
\]

Define
\[
\CCC(\HH) = \corel_{\HH} = (\corel_{\HH}^{i})_{i \in P^{\bA}}.
\]

\begin{lem}\label{lem-cccmor}
$\cormap{\CCC(\HH)}{\CCC(\bA)}{\CCC(\bA')}$ such that:
\begin{enumerate}

\item\label{lem-cccmor-1}
For $p \in P^{\bA}$, $\Gamma \in \LLL(P^{\bA})$ and $\varphi \in \LLL(P^{\bA'})$,
\[ 
\Gamma \corel_{\HH}^{p} \varphi \Rightarrow \ov{\Gamma} \in \con_{p} \kand \ov{\Gamma} H_{p} \ov{\varphi}.
\]

\item\label{lem-cccmor-2}
For $i \in A$, $X \in \con_{i}$ and $Y \fsubset A'$, 
\[
X H_{i} Y \Rightarrow X \in \powf{\LLL(P^{\bA}_{i})} \cup \{\{i\}\} \kand X \corel_{\HH}^{i} \bigwedge Y.
\]
\end{enumerate}
\end{lem}
\begin{pf}
We have to verify the conditions in Definition~\ref{dn-glob}. 

For Condition~\ref{dn-glob}\eqref{dn-glob-1}  we have to show closure under the Rules (RT), (W), (Lcut), (Rcut): (L$\land$) and (R$\land$). (RT) follows with Property~\ref{dn-am}\eqref{dn-am-2}  from the fact that $\HH$ respects truth elements. The closure under (W) follows as in the proof of Theorem~\ref{thm-frlog}, but now with Property~\ref{dn-am}\eqref{dn-am-2}; the closure under (Lcut) is a consequence of \ref{dn-am}\eqref{dn-am-3} and the closure under (Rcut) of \ref{dn-am}\eqref{dn-am-1}. The closure under (L$\land$) and (R$\land$) follows as in the proof of Theorem~\ref{thm-frlog}.

Conditions~\ref{dn-glob}\eqref{dn-glob-2} and \ref{dn-glob} \eqref{dn-glob-3} follow with \ref{dn-am}\eqref{dn-am-4} and \ref{dn-am}\eqref{dn-am-5}, respectively.

Statements~\eqref{lem-cccmor-1} and \eqref{lem-cccmor-2}  hold by definition and \eqref{eq-logfr}.
\end{pf}

\begin{pn}\label{pn-ccc-fctor}
$\fun{\CCC}{\SIF_{\bt}}{\CSL}$ is a faithful functor.
\end{pn}

\begin{thm}\label{thm-embed}
$\CCC$ is an embedding of $\SIF_{\bt}$ into $\CSL$.
\end{thm}

Next, we will examine the reverse situation, namely how a continuous stratified conjunctive logic determines a strong continuous information frame. Let $\PP$ be a continuous stratified conjunctive logic. Set $A_{\PP} = P$, $\bt_{\PP} = \top$,
\begin{gather*}
\con^{\PP}_{p} = \{\{p\}\} \cup \set{X \fsubset P_{p}}{p \vdash^{p} \bigwedge X},  \\
X \vDash^{\PP}_{p} q \Leftrightarrow \bigwedge X \vdash^{p} q, 
\end{gather*}
with $p, q \in P$ and $X \in \con^{\PP}_{p}$.

\begin{lem}\label{lem-logifr}
Let $\PP$ be a continuous stratified conjunctive logic. Then 
\[
\bA_{\PP} =  (A_{\PP}, (\con^{\PP}_{i})_{i \in A_{\PP}}, (\vDash^{\PP}_{i})_{i \in A_{\PP}})
\]
 is a strong continuous information frame with a truth element, such that for all $p \in P$ the following statements hold:
\begin{enumerate}

\item\label{lem-logifr-0}
For all $X \fsubset A_{\PP}$,
\[
X \in \con^{\PP}_{i} \Leftrightarrow \bigwedge X \in \powf{\LLL(P_{p})} \cup \{\{p\}\}.
\]

\item\label{lem-logifr-1}
For all $X \in \con^{\PP}_{p}$ and $Y \fsubset P$, 
\[
X \vDash^{\PP}_{p} Y \Rightarrow X \vdash^{p} \bigwedge Y.
\]

\item\label{lem-logifr-2}
For all $\Gamma \in \powf{\LLL(P_{p})} \cup \{\{p\}\}$ and $\varphi \in \LLL(P_{p})$,
\[
\Gamma \vdash^{p} \varphi \Rightarrow \ov{\Gamma} \vDash^{\PP}_{p} \ov{\varphi}.
\]
\end{enumerate}
\end{lem}
\begin{pf}
We first show that the conditions in Definition~\ref{dn-infofr} hold:

Condition~\ref{dn-infofr}\eqref{dn-infofr-1} holds by definition. For Condition~\ref{dn-infofr}\eqref{dn-infofr-2} let  $p \in A_{\PP}$, $X \in \con^{\PP}_{p}$ and $Y \subseteq X$. If $X = \{p\}$ then $Y = X$ or $Y= \emptyset$. Without restriction we only consider the second case. Then $\bigwedge Y = \top$.  Because of (R$\top$) we have that $p \vdash^{p} \top$. Hence, $Y \in \con^{\PP}_{p}$. If $X \neq \{p\}$, then $p \vdash^{p} \bigwedge X$, from which it follows with (R$\land$) that also $p \vdash^{p} \bigwedge Y$. Thus, in this case, $Y \in \con^{\PP}_{p}$ also applies.

For Condition~\ref{dn-infofr}\eqref{dn-infofr-4} assume that $X \in \con^{\PP}_{p}$ and $Y \fsubset A_{\PP}$ with $X \vDash^{\PP}_{p} Y$. If $X \neq \{p\}$, we have that $p \vdash^{p} \bigwedge X$ and $\bigwedge X \vdash^{p} \bigwedge Y$. Using (Cut) it follows that $p \vdash^{p} \bigwedge Y$, which we also have when $X = \{p\}$. So, it follows that $Y \in \con^{\PP}_{p}$. 

For \ref{dn-infofr}\eqref{dn-infofr-5} let $q \in A_{\PP}$ and $X, Y \in \con^{\PP}_{p}$ with $X \subseteq Y$ and $X \vDash^{\PP}_{p} q$. Then we have that $\bigwedge X \vdash^{p} q$. 
Without restriction assume that $X \neq Y$. Suppose first that $Y = \{p\}$. Then $X \neq \{p\}$ and therefore $p \vdash^{p} \bigwedge X$. It follows that $p \vdash^{p} q$. Thus, $Y \vdash^{p} q$ if $Y = \{p\}$. If $Y \neq \{p\}$, we have for $Z = Y \setminus X$ that $\bigwedge Z \in \LLL(P_{q})$. With Rule (W) we therefore obtain that $\bigwedge X, \bigwedge Z \vdash^{p} q$, from which it follows with Lemma~\ref{lem-rulebarwedge} that also $\bigwedge Y \vdash^{p} q$. So, we have in both cases that $Y \vDash^{\PP}_{p} q$.

Condition~\ref{dn-infofr}\eqref{dn-infofr-6} is a consequence of (Cut). For Condition~\ref{dn-infofr}\eqref{dn-infofr-7} let $p, q \in A_{\PP}$ with $\{p\} \in \con^{\PP}_{q}$. Without restriction assume that $p \neq q$. Then $q \vdash^{q} p$. Let $X \in \con^{\PP}_{p}$ and without restriction suppose that $X \neq \{p\}$. Then we have that $p \vdash^{p}\bigwedge X$ and because of Property~\ref{dn-stratlog}\eqref{dn-stratlog-2} also that $p \vdash^{q}\bigwedge X$. It follows that $q \vdash^{q} \bigwedge X$. Thus, $X \in \con^{\PP}_{q}$.
Condition~\ref{dn-infofr}\eqref{dn-infofr-9} is a direct consequence of Property~\ref{dn-stratlog}\eqref{dn-stratlog-2}. 

It remains to consider Condition~\ref{dn-infofr}\eqref{dn-infofr-11}. Assume that $X \vDash^{\PP}_{p} Y$. Then we have that $\bigwedge X \vdash^{p} \bigwedge Y$. Because of Property (INT) there are $q \in P_{p}$ and $\psi \in \LLL(P_{q})$ with $\bigwedge X \vdash^{p} q \land \psi$ and $\psi \vdash^{q} \bigwedge Y$. Since $\psi \in \LLL(P_{q})$, we have that $q \vdash^{q} \bigwedge \ov{\psi}$. Thus, $\ov{\psi} \in \con^{\PP}_{q}$.  Moreover, we have that $X \vDash^{\PP}_{p} \{q\} \cup \ov{\psi}$ and $\ov{\psi} \vDash^{\PP}_{q} Y$. Set $Z = \ov{\psi}$.

This shows that $\bA_{\PP}$ is a continuous information frame. To show that $\bA_{\PP}$ is also strong, let $p \in A_{\PP}$ and $X \in \con^{\PP}_{p}$ with $X \neq \{p\}$. Then $p \vdash^{p} \bigwedge X$ and hence, $\{p\} \vDash^{\PP}_{p} X$, With (R$\top$) it finally follows that $\top$ is a truth element.

It remains to prove Statements~\eqref{lem-logifr-0}-\eqref{lem-logifr-2}. Statement~\eqref{lem-logifr-0} is a consequence of Condition~\eqref{cn-S} and Statement~\eqref{lem-logifr-1}  follows with (R$\land$) from the definition of the entailment relation. For Statement~\eqref{lem-logifr-2} assume that $\Gamma \vdash_{p} \varphi$. With Lemma~\ref{lem-rulebarwedge}  and (R$\land$) it follows that for all $q \in \ov{\varphi}$, $\bigwedge \ov{\Gamma} \vdash^{p} q$ and hence $\ov{\Gamma} \vDash^{\PP}_{p} q$. Thus, we have that $\ov{\Gamma} \vDash^{\PP}_{p} \ov{\varphi}$.
\end{pf}

Let $\PP'$ be a further continuous stratified conjunctive logic and $\cormap{\corel}{\PP}{\PP'}$. For $p \in A_{\PP}$ define $H^{\corel}_{p} \subseteq \con^{\PP}_{p} \times P'$ by
\[
X H^{\corel}_{p} q \Leftrightarrow \bigwedge X \corel^{p} q.
\]
and set $\HH_{\corel} = (H^{\corel}_{p})_{p \in A_{\PP}}$.

\begin{lem}\label{pn-corelapp}
Let $\PP$ and $\PP'$ be continuous stratified conjunctive logics and 
\[\cormap{\corel}{\PP}{\PP'}.
\]
 Then the following statements hold:
\begin{enumerate}

\item\label{pn-corelapp-1}
$\appmapf{\HH_{\corel}}{\bA_{\PP}}{\bA_{\PP'}}$.

\item\label{pn-corelapp-2}
$\HH_{\corel}$ respects  truth elements.

\item\label{pn-corelapp-3}
For all $p \in A_{\PP}$, $X \in \con^{\PP}_{p}$ and $Y \fsubset A_{\PP'}$,
\[
X H^{\corel}_{p} Y \Rightarrow \bigwedge X \corel^{p} \bigwedge Y.
\]

\item\label{pn-corelapp-4}
For all $p \in P$, $\Gamma \in \powf{\LLL(P_{p})} \cup \{\{p\}\}$ and $\varphi \in \LLL(P')$,
\[
\Gamma \corel^{p} \varphi \Rightarrow \ov{\Gamma} H^{\corel}_{p} \ov{\varphi}.
\]

\end{enumerate}
\end{lem}
\begin{pf}
For Statement~\eqref{pn-corelapp-1} we have to show that the conditions in Definition~\ref{dn-am} hold. For Condition~\ref{dn-am}\eqref{dn-am-1} assume that $X H^{\corel}_{p} (\{r\} \cup Y)$ and $Y \vDash^{\PP'}_{r} q$. Then $\bigwedge X \corel^{p} r \land \bigwedge Y$ and $\bigwedge Y \vdash'^{r} q$.  Hence, $\bigwedge X \corel^{p} q$, using (Rcut). Thus, we have that $X H^{\corel}_{p} q$.

For Condition~\ref{dn-am}\eqref{dn-am-2} suppose that $X, Y \in \con^{\PP}_{p}$ with $X \subseteq Y$ so that $X H^{\corel}_{p} q$. Without restriction let $X \neq Y$. If $Y = \{p\}$, then  $X \neq \{p\}$ and therefore $p \vdash^{p} \bigwedge X$. With (Lcut) it follows that $p \corel^{p} q$, that is,  we have $Y H^{\corel}_{p} q$.  If $Y \neq \{p\}$, we have for $Z = Y \setminus X$ that $\bigwedge Z \in \LLL(P_{q})$. With Rule (W) we therefore obtain that $\bigwedge X, \bigwedge Z \corel^{p} q$, from which it follows that $Y \corel^{p} q$. So, we have in both cases that $Y H^{\corel}_{p} q$.

Condition~\ref{dn-am}\eqref{dn-am-3} is a consequence of (Lcut) and Condition~\ref{dn-am}\eqref{dn-am-4} follows with Condition~\ref{dn-glob}\eqref{dn-glob-2}. For Condition~\ref{dn-am}\eqref{dn-am-5} assume that $X H^{\corel}_{p} F$. Then $\bigwedge X \corel^{p} \bigwedge F$. Because of Condition~\ref{dn-glob}\eqref{dn-glob-3} there are $r \in P$, $\theta \in \LLL(P_{r}) \cup \{r\}$, $q \in P'$ and $\psi \in \LLL(P'_{q}) \cup \{q\}$ so that $\bigwedge X \vdash^{p} r \land \theta$, $\theta \corel^{r} q \land \psi$ and $\psi \vdash'_{q} \bigwedge F$. Set $U = \ov{\theta}$ and $V = \ov{\psi}$. Then $U \in \con^{\PP}_{r}$ and $V \in \con^{\PP'}_{q}$. Moreover, $X \vDash^{\corel}_{p} (\{r\} \cup U)$, $U H^{\corel}_{r} (\{q\} \cup V)$ and $V \vDash^{\PP'}_{q} F$.

This shows that $\HH^{\corel}$ is an approximating family. With Rule (R$\top$) it follows that $\HH^{\corel}$ respects truth elements, that is Statement~\eqref{pn-corelapp-2} holds. Statements~\eqref{pn-corelapp-3} and \eqref{pn-corelapp-4} are immediate consequences of the definition. Only  Rules~\eqref{rule-bar} and (R$\land$) are applied.
\end{pf}

\begin{pn}\label{pn-eeefctor}
Let $\PP$ and $\PP'$ be continuous stratified conjunctive logics and $\cormap{\corel}{\PP}{\PP'}$. Set
\[
\EEE(\PP) = \bA_{\PP} \qquad\text{and}\qquad \EEE(\corel) = \HH_{\corel}.
\]
Then $\fun{\EEE}{\CSL}{\SIF_{\bt}}$ is a functor.
\end{pn}

In the rest of this section we will show that the functors $\CCC$ and $\EEE$ are an isomorphism between the categories $\CSL$ and $\SIF_{\bt}$. 

Let $\bA$ be a strong continuous information frame. Then we have that
\[
\EEE(\CCC(\bA)) = \EEE(\PP^{\bA}) = \bA_{\PP^{\bA}} = (A_{\PP^{\bA}}, (\con^{\PP^{\bA}}_{i})_{i \in A_{\PP^{\bA}}}, (\vDash^{\PP^{\bA}}_{i})_{i \in A_{\PP^{\bA}}}),
\]
where
\begin{gather*}
A_{\PP^{\bA}} = P^{\bA} = A, \qquad \bt_{\PP^{\bA}}  = \top_{\bA} = \bt, \\
X \in \con^{\PP^{\bA}}_{i} \Leftrightarrow \bigwedge X \in \LLL(P^{\bA}_{i}) \cup \{\{i\}\} \Leftrightarrow \ov{\bigwedge X} \in \con_{i} \Leftrightarrow X \in \con_{i}, \\
X \vDash^{\PP^{\bA}}_{i} j \Leftrightarrow X \vdash_{\bA}^{i} j \Leftrightarrow X \vDash_{i} j.
\end{gather*}

Let $\bA'$ be a further strong continuous information frame and $\appmapf{\HH}{\bA}{\bA'}$. Then we have in addition that
\[
X H^{\corel_{\HH}}_{i} j \Leftrightarrow X \corel_{\HH}^{i} j  \Leftrightarrow X H_{i} j.
\]

\begin{pn}\label{pn-eeeccc}
$\EEE \circ \CCC = \ID_{\SIF_{\bt}}$.
\end{pn}

\begin{cor}\label{cor-eeefull}
$\fun{\EEE}{\CSL}{\SIF_{\bt}}$ is full.
\end{cor}

Next, we  consider the reverse situation.  Let to this end $\PP$ be a continuous stratified conjunctive logic. Then we have that
\[
\CCC(\EEE(\PP)) = \CCC(\bA_{\PP}) = \PP^{\bA_{\PP}} = (P^{\bA_{\PP}}, (P^{\bA_{\PP}}_{p})_{p \in P^{\bA_{\PP}}}, (\vdash^{p}_{\bA_{\PP}})_{p \in P^{\bA_{\PP}}}),
\] 
where
\[
P^{\bA_{\PP}} = A_{\PP} = P, \qquad \top^{\bA_{\PP}} = \bt_{\PP} = \top;
\]
moreover, for $p, q \in P$,
\[
q \in P^{\bA_{\PP}}_{p} 
\Leftrightarrow p \vdash^{p}_{\bA_{\PP}} q
\Leftrightarrow \{p\} \vDash_{p}^{\PP} q
\Leftrightarrow p \vdash^{p} q,
\]
and for $\Gamma \in \powf{\LLL(P^{\bA_{\PP}}_{p})} \cup \{\{p\}\}$ and $\varphi \in \LLL(P^{\bA_{\PP}}_{p})$,
\begin{align*}
\Gamma \vdash_{\bA_{\PP}}^{p} \varphi 
\Leftrightarrow \ov{\Gamma} \vDash^{\PP}_{p} \ov{\varphi} 
&\Leftrightarrow (\forall q \in \ov{\varphi})\, \ov{\Gamma} \vDash^{\PP}_{p} q \\
&\Leftrightarrow  (\forall q \in \ov{\varphi})\, \bigwedge \ov{\Gamma} \vdash^{p} q
\Leftrightarrow \bigwedge \ov{\Gamma} \vdash^{p} \varphi
\Leftrightarrow \Gamma \vdash^{p} \varphi.
\end{align*}
Here, the last equivalence follows with Lemma~\ref{lem-rulebarwedge}.

Let $\PP'$ be a further continuous stratified conjunctive logic and $\cormap{\corel}{\PP}{\PP'}$. Then if follows similarly that
\[
\Gamma \corel_{\HH^{\corel}}^{p} \varphi \Leftrightarrow \Gamma \corel^{p} \varphi.
\]

\begin{pn}\label{pn-isoccceee}
$\CCC \circ \EEE = \ID_{\CSL}$.
\end{pn}

\begin{thm}\label{thm-isocslsift}
The categories $\CSL$ and $\SIF_{\bt}$ are isomorphic.
\end{thm}

\begin{cor}\label{cor-eqcscif}
The categories $\CSL$, $\SIF_{\bt}$, $\SIF$, $\CIF$, $\CIF_{\bt}$, $\CIS$ and $\SCIS$ are all equivalent.
\end{cor}

\section{Conclusion}\label{sec-concl}

The study of representations of domains by other than order structures is an important branch of domain theory. Such representations often provide an easily understandable approach to domain theory. The wide variety of such approaches also highlights the relationship of domain theory to other areas of mathematics and underscores its fundamental importance.

One of the first such representations was presented by D.\ Scott~\cite{sc82} in a seminal paper in 1982. In it, he introduced information systems and showed that they exactly represent bounded-complete algebraic domains. This work paved the way for a large number of similar representations of various subclasses of domains. One such representation for the general class of all continuous domains was presented in \cite{sxx08}. The introduction of information frames instead of information systems was motivated by the goal of visualising consistency witnesses that confirm the consistency of information associated with finite token sets in general information systems representing continuous domains. This is particularly the case for strong information frames.

In the present paper it is shown that the categories of continuous information systems as introduced in~\cite{sxx08}, the category of continuous information frames as studied in~\cite{sp25} and the present paper, and the subcategory of strong continuous information frames are all equivalent, which in particular answers a question of Prof.\ Luoshan Xu at \emph{ISDT 2024}.

Continuous information frames are families of rudimentary logics, each member of which possesses a consistency predicate and an entailment relation. However, they lack the expressive power of propositional formulas. In a first attempt to investigate the relationship between continuous information frames---and thus continuous domains---and propositional logic, each of these rudimentary logics is extended into a conjunctive fragment of propositional logic. It turns out that the category of families of such conjunctive fragments is isomorphic to the category of continuous information frames with a truth element. This shows, in particular, that the expressive power of propositional logic results from the interplay of the two propositional connectives, conjunction and disjunction.

\section*{Acknowledgement}

The author thanks Prof. Luoshan Xu for stimulating discussions. Thanks also go to the anonymous reviewer for careful reading of the manuscript and helpful comments.

\end{document}